\documentclass[12pt]{article}
\usepackage[latin1]{inputenc} 
\usepackage{amssymb}
\usepackage{amsfonts} 
\usepackage{amsthm} 
\usepackage{amsmath} 
\usepackage{hyperref}

\def\ni{\noindent}
\def\nn{\nonumber}

\def \bc {\begin{center}}
\def \ec {\end{center}}
\def \bi {\begin{itemize}}
\def \ei {\end{itemize}}

\def \ba {\begin{array}}
\def \ea {\end{array}}

\def \bea {\begin{eqnarray}}
\def \eea {\end{eqnarray}}

\def \be {\begin{equation}}
\def \ee {\end{equation}}

\def \lb {\left[}
\def \rb {\right]}
\newcommand{\la}{\langle}
\newcommand{\ra}{\rangle}

\def \um {\frac{1}{2}}

\def\tr {{\rm tr}}

\def\cD {{\cal D}}
\def\cL {{\cal L}}

\def\cU {{\cal U}}

\def\cS {{\cal S}}

\def\mbN {{\mathbb N}}
\def\mbD {{\mathbb D}}

\def\mbC {{\mathbb C}}
\def\mbT {{\mathbb T}}

\newcommand{\bra}[1]{ \left<#1\right| }
\newcommand{\ket}[1]{ \left|#1\right> }
\newcommand{\scprod}[2]{ \left<#1\right.|\left.#2\right> }
\newcommand{\brap}[1]{ \left(#1\right| }
\newcommand{\ketp}[1]{ \left|#1\right)}
\newcommand{\scprodp}[2]{ \left(#1\right.|\left.#2\right) }

\newtheorem{thm}{Theorem}[section]

\newtheorem{prop}[thm]{Proposition}

\theoremstyle{remark}
\newtheorem{rem}[thm]{Remark}

\begin{document}

\begin{center}
{\LARGE {\bf Conformal Spinning Quantum Particles in Complex
Minkowski Space as Constrained Nonlinear Sigma Models in $U(2,2)$ and
Born's Reciprocity}}
\end{center}
\bigskip
\bigskip

\centerline{{\sc M. Calixto}$^{1,2}$\footnote{Corresponding author:
Manuel.Calixto@upct.es} and {\sc E. Pérez-Romero}$^{2}$ }

\bigskip

\bc {\it $^1$ Departamento de Matemática Aplicada y Estad\'\i stica,
Universidad Politécnica de Cartagena, Paseo Alfonso XIII 56, 30203
Cartagena, Spain}
\\
{\it $^2$ Instituto de Astrof\'\i sica de Andaluc\'\i a
(IAA-CSIC), Apartado Postal 3004, 18080 Granada, Spain}

 \ec

\bigskip
\begin{center}
{\bf Abstract}
\end{center}
\small
\begin{list}{}{\setlength{\leftmargin}{3pc}\setlength{\rightmargin}{3pc}}
\item We revise the use of 8-dimensional conformal, complex
(Cartan) domains as a base for the construction of conformally
invariant quantum (field) theory, either as phase or configuration
spaces. We follow a gauge-invariant Lagrangian approach (of
nonlinear sigma-model type) and use a generalized Dirac method for
the quantization of constrained systems, which resembles in some
aspects the standard approach to quantizing coadjoint orbits of a
group $G$. Physical wave functions, Haar measures, orthonormal
basis and reproducing (Bergman) kernels are explicitly calculated
in and holomorphic picture in these Cartan domains for both scalar
and spinning quantum particles. Similarities and differences with
other results in the literature are also discussed and an
extension of Schwinger's Master Theorem is commented in connection
with closure relations. An adaptation of the Born's Reciprocity
Principle (BRP) to the conformal relativity, the replacement of
space-time by the 8-dimensional conformal domain at short
distances and the existence of a maximal acceleration are also put
forward.
\end{list}
\normalsize 





\newpage
\section{Introduction}

Complex manifolds and, in particular, Cartan classical domains have been
studied for many years by mathematicians and theoretical
physicists (see e.g. \cite{Coquereaux} and references therein for a review).
In this article we are interested in the Lie ball
\[\mbD=SO(4,2)/(SO(4)\times SO(2))=SU(2,2)/S(U(2)\times U(2)),\]
which can be mapped one-to-one onto the 8-dimensional forward/future tube domain
\[\mbT=\{x^\mu+iy^\mu\in\mbC^{1,3},\;\;y^0>\|\vec y\|\}\]
of the complex Minkowski space $\mbC^{1,3}$ through a Cayley
transformation (see next Section for more details). Both manifolds
can be considered as the phase space of massive conformal
particles and there is a renewed interest in its quantization (see
e.g. \cite{Odzijewicz} and references therein for a survey). The
presentation followed in the literature is of geometric (twistor
\cite{Penrose,Penrose2} and Konstant-Kirillov-Souriau
\cite{Souriau,Kirillov} descriptions) and representation-theoretic
\cite{Ruhl0,Ruhl1} nature. Here we shall adopt a
(sigma-model-type) Lagrangian approach to the subject and we shall
use a generalized Dirac method for the quantization of constrained
systems which resembles in some aspects the particular approach to
quantizing coadjoint orbits of a group $G$ developed many years
ago in \cite{Bal} (see also \cite{Bal2} and \cite{hamilredu} for
interesting examples in $G=SU(3)$).

We share with many authors (namely,
\cite{Coquereaux,Odzijewicz,KaiserAP,Kaiser,Jadczyk,Born1,Born2,Castro0,Castro1,Caianiello,Castro2,Low1,Low2}) the
believe that the use of complex Minkowski 8-dimensional space as a
base for the construction of quantum (field) theory is not only
useful from the technical point of view but can be of great
physical importance. Actually, as suggested in \cite{Jadczyk}, the
conformal domain $\mathbb D$ could be considered as the
replacement of the space-time at short distances (at the
``microscale''). This interpretation is based on Born's
Reciprocity Principle (BRP) \cite{Born1,Born2}, originally
intended to merging quantum theory and general relativity. The
reciprocity symmetry between coordinates $x_\mu$ and momenta
$p_\mu$ states that the laws of nature are (or should be)
invariant under the transformations
\be(x_\mu, p_\mu)\to (\pm p_\mu,\mp x_\mu).\label{BRP}\ee
The word ``reciprocity'' is used in analogy with the
lattice theory of crystals, where some physical phenomena (like
the theory of diffraction) are sometimes better described in the
$p$-space  by means of the reciprocal (Bravais) lattice. The
argument here is that Born's reciprocity implies that there must
be a reciprocally conjugate relativity principle according to
which the rate of change of momentum (force) should be bounded by
a universal constant $b$, much in the same way the usual
relativity principle implies a bound of the rate of change of
position (velocity) by the speed of light $c$. As a consequence of
the BRP, there must exist a minimum (namely, Planck) length
$\ell_{\rm min}=\sqrt{\hbar c/b}$.

This symmetry led Born to conjecture that the basic underlying
physical space is the 8-dimensional $\{x_\mu,p_\mu\}$ and to
replace the Poincaré invariant line element $d\tau^2=dx_\mu
dx^\mu$ by the Finslerian-like metric (see \cite{Castro0,Castro1}
for an extension to Born-Clifford phase spaces)
\be d\tilde\tau^2=dx_\mu dx^\mu+\frac{\ell^4_{\rm
min}}{\hbar^2}dp_\mu dp^\mu.\label{Bornline}\ee
From the BRP point of view, local (versus extended) field theories like
Klein-Gordon's represent the ``point-particle limit'' $\ell_{\rm min}\to
0$, for which the reciprocal symmetry is broken. Also, the Minkowski
spacetime is interpreted either as a local ($\ell_{\rm min}\to 0$) version
or as a high-energy-momentum-transfer limit ($b\to\infty$) of this
8-dimensional phase-space domain. Moreover, putting $dp_\mu/d\tau=m
d^2x_\mu/d\tau^2=ma_\mu$, with $m=b\ell_{\rm min}/c^2$ a (namely, Planck)
mass and $a_\mu$ the proper acceleration (with $a^2\leq 0$, space-like),
one can write the previous extended line element as
\be d\tilde\tau=d\tau\sqrt{1-\frac{|a^2|}{a^2_{\rm
max}}},\label{amaxBorn}\ee
which naturally leads to a \textit{maximal (proper) acceleration}
$a_{\rm max}=c^2/\ell_{\rm min}$. The existence and physical
consequences of a maximal acceleration was already derived by
Caianiello \cite{Caianiello}. Many papers have been published in
the last years (see e.g. \cite{Castro2} and references therein),
each one introducing the maximal acceleration starting from
different motivations and from different theoretical schemes.
Among the large list of physical applications of Caianiello's
model we would like to point out the one in cosmology which avoids
an initial singularity while preserving inflation. Also, a
maximal-acceleration relativity principle leads to a variable fine
structure ``constant'' $\alpha$ \cite{Castro2}, according to which
$\alpha$ could have been extremely small (zero) in the early
Universe and then all matter in the Universe could have emerged
via the Fulling-Davies-Unruh-Hawking effect (vacuum radiation due to the
acceleration with respect to the vacuum frame of reference)
\cite{Fulling,Davies,Unruh,Hawking}.

There has been group-theoretical revisions of the BRP like
\cite{Low1,Low2} replacing the Poincaré by the Canonical (or
Quaplectic) group of reciprocal relativity, which enjoys a richer
structure than Poincaré. In this article we pursue a different
reformulation of BRP as a natural symmetry inside the conformal
group $SO(4,2)$ and the replacement of space-time by the
8-dimensional conformal domain $\mathbb D$ or $\mathbb T$ at short
distances. We believe that new interesting physical phenomena
remain to be unravelled inside this framework. Actually, in a
coming paper \cite{Unruhconf} (see also \cite{conforme} for a
previous related work), we shall discuss a group-theoretical
revision of the Unruh effect  \cite{Unruh} as a spontaneous
breakdown of the conformal symmetry and the consequences of a
maximal acceleration. Also, a wavelet transform on the tube domain
$\mathbb T$, based on the conformal group, could provide a way to
analyze wave packets localized in both: space and time. Important
developments in this direction have been done in
\cite{Kaiser1,Kaiser2} for electromagnetic (massless) signals and
\cite{EMSMTA} for fields with continuous mass spectrum.

In this article we shall study the geometrical and quantum mechanical underlying framework.
We shall follow a gauge-invariant
(singular) Lagrangian approach of nonlinear sigma-model type and we shall use a generalized
Dirac method for the quantization of constrained systems.

The paper is organized as follows. In Section \ref{confcoord} we
briefly review the conformal group $SO(4,2)\simeq SU(2,2)$, its
Lie algebra generators and commutators, and provide different
coordinate systems for the conformal domains $\mathbb D$ and
$\mathbb T$; in this Section we also introduce the concept of BRP
in a conformally invariant setting. Section \ref{NLSMsec} is devoted
to the Lagrangian formulation of conformally invariant nonlinear
sigma-models on the conformal domains (either as configuration or
phase spaces) and the study of their gauge invariance. The
quantization of these models (for the case of Lagrangians linear
in velocities) is accomplished in Section \ref{quantizationsec} by
using a generalized Dirac method for the quantization of
constrained systems which resembles in some aspects the particular
approach to quantizing coadjoint orbits of $G$. Physical wave
functions, Haar measures, orthonormal basis and
reproducing (Bergman) kernels are explicitly calculated  in an
holomorphic picture in the Cartan domain $\mathbb D$, for both
scalar and spinning quantum particles in subsections
\ref{scalarsec} and \ref{spinningsec}, respectively. Similarities
and differences with other results in the literature are also
discussed and an extension of the Schwinger Master Theorem is
commented in connection with closure relations. In Section
\ref{tubedomainsec} we translate (through an equivariant map) all
the constructions above to the tube domain $\mathbb T$, where we
enjoy more physical intuition. We comment on K\"ahler structures
and generalized Born-like line elements and the existence of a
maximal acceleration for conformal (quantum) particles. The last
Section \ref{comments} is devoted to comments and outlook where we
point out an interesting connection between BRP and CPT symmetry
inside the conformal group and discuss on the appearance of a
maximal acceleration in this scheme.

\section{\label{confcoord}The conformal symmetry in 1+3D: coordinate systems and generators}

The conformal group $SO(4,2)$ is comprised of Poincaré (spacetime
translations $b^\mu\in\mathbb R^{1,3}$ and Lorentz $\Lambda^{\mu}_\nu\in
SO(3,1)$) transformations augmented by dilations ($\rho=e^\tau\in\mathbb
R_+$) and relativistic uniform accelerations (special conformal
transformations, SCT,  $a^\mu\in\mathbb R^{1,3}$) which, in Minkowski
spacetime, have the following realization:
\be\ba{ll} x'^\mu = x^\mu+b^\mu, & x'^\mu=\Lambda^{\mu}_\nu(\omega) x^\nu,\\
x'^\mu=\rho x^\mu, &x'^\mu=\frac{x^\mu+a^\mu x^2}{1+2a x+a^2
x^2},\ea\label{confact} \ee
respectively. The interpretation of SCT as
transitions from inertial reference frames to systems of relativistic,
uniformly accelerated observers was identified many years ago by (see
e.g.,  \cite{Hill,conforme-ac,Cervero1975}), although alternative meanings
have also been proposed. One is related to the Weyl's idea of different
lengths in different points of space time \cite{Weyl}: ``the rule for
measuring distances changes at different positions''. Other is Kastrup's
interpretation of SCT as geometrical gauge transformations of the
Minkowski space \cite{Kastrup1} (for this point see later on Eq.
(\ref{constrained4})).

The generators of the transformations (\ref{confact}) are easily deduced:
\be \ba{rcl}P_\mu &=& \frac{\partial}{\partial x^\mu}, \;\;
M_{\mu\nu}=x_\mu
\frac{\partial}{\partial x^\nu}-x_\nu \frac{\partial}{\partial x^\mu},\\
D&=&x^\mu\frac{\partial}{\partial x^\mu},\;\; K_\mu=-2x_\mu x^\nu
\frac{\partial}{\partial x^\nu}+x^2\frac{\partial}{\partial x^\mu} \label{confvf}\ea\ee
and they close into the conformal Lie algebra
 \be\ba{rcl} \lb M_{\mu\nu},M_{\rho\sigma}\rb &=&\eta_{\nu\rho}M_{\mu\sigma}+\eta_{\mu\sigma}M_{\nu\rho}
-\eta_{\mu\rho}M_{\nu\sigma}-\eta_{\nu\sigma}M_{\mu\rho},\\
\left[P_\mu,M_{\rho\sigma}\right] &=& \eta_{\mu\rho} P_\sigma -
\eta_{\mu\sigma}
P_\rho,\;\; \lb P_\mu,P_\nu\rb=0,\\
\lb K_\mu,M_{\rho\sigma}\rb &=& \eta_{\mu\rho}K_\sigma-\eta_{\mu\sigma}K_\rho,\;\; \lb K_\mu,K_\nu\rb=0, \\
\lb D,P_\mu\rb &=&-P_\mu, \;\;\lb D,K_\mu\rb =K_\mu,\;\; \lb D,M_{\mu\nu}\rb=0,\\
\lb K_\mu,P_\nu\rb &=& 2(\eta_{\mu\nu}
D+M_{\mu\nu}).\ea\label{conformalgebra}\ee
We shall argue later that $P_\mu$ and $K_\mu$ are conjugated variables
(they can not be simultaneously measured) and that $D$ can be taken to be
the generator of (proper) time translations (i.e., the Hamiltonian). A
BRP-like symmetry manifests here in the form: \be P_\mu\to K_\mu,\;
K_\mu\to P_\mu,\; D\to -D,\label{BRPconf}\ee
which leaves the commutation relations (\ref{conformalgebra}) unaltered. This symmetry can also be seen in the
quadratic Casimir operator:
%
\be C_2=D^2-\um M_{\mu\nu}M^{\mu\nu}+\um(P_\mu K^\mu+K_\mu P^\mu)
=D^2-\um M_{\mu\nu}M^{\mu\nu}+P_\mu K^\mu +4 D,\label{Casimir}\ee
which generalizes the Poincaré Casimir $P^2=P_\mu P^\mu$, just as
$d\tilde\tau$ in (\ref{Bornline}) generalizes the Poincaré
invariant line element $d\tau$. We shall provide a conformal
invariant line element similar to $d\tilde\tau$ later in Section
\ref{tubedomainsec}.

Any group element $g\in SO(4,2)$ (near the identity element $1$) could be
written  as the exponential map
\be g=\exp(u),\; u=\tau D+b^\mu P_\mu+a^\mu
K_\mu+\omega^{\mu\nu}M_{\mu\nu},\label{expmap}\ee
of the Lie-algebra element $u$ (see \cite{expomap0,expomap1}). The compactified
Minkowski space $\mathbb M=\mathbb S^3\times_{\mathbb Z_2} \mathbb
S^1\simeq U(2)$ can be obtained as the coset $\mathbb M=SO(4,2)/\mathbb W$,
where $\mathbb W$ denotes the Weyl subgroup generated by $K_\mu,
M_{\mu\nu}$ and $D$ (i.e., a Poincaré subgroup $\mathbb P=SO(3,1)\circledS
\mathbb R^4$ augmented by dilations $\mathbb R^+$). The Weyl group
$\mathbb W$ is the stability subgroup (the little group in physical usage)
of $x^\mu=0$.

There is another interesting realization of the conformal Lie algebra
(\ref{conformalgebra}) in terms of gamma matrices in, for instance, the
Weyl basis
\be \gamma^\mu=\left(\ba{cc} 0& \sigma^\mu \\ \check{\sigma}^\mu
&0\ea\right),\;\;
\gamma^5=i\gamma^0\gamma^1\gamma^2\gamma^3=\left(\ba{cc}
-\sigma^0& 0\\ 0& \sigma^0\ea\right),\nn\ee
where $\check{\sigma}^\mu\equiv \sigma_\mu$ (we are using the
convention $\eta={\rm diag}(1,-1,-1,-1)$ for the Minkowski metric)
and $\sigma^\mu$ are the Pauli matrices
\be \sigma^0=\left(\ba{cc} 1& 0
\\ 0 &1\ea\right),\;\sigma^1=\left(\ba{cc} 0& 1
\\ 1 &0\ea\right),\;\sigma^2=\left(\ba{cc} 0& -i
\\ i &0\ea\right),\;\sigma^3=\left(\ba{cc} 1& 0
\\ 0 &-1\ea\right).\nn\ee
Indeed, the choice
\be\ba{rcl} D&=&\frac{\gamma^5}{2},\;M^{\mu\nu}=\frac{\lb
\gamma^\mu,\gamma^\nu\rb}{4}=\frac{1}{4}\left(\ba{cc} \sigma^\mu\check{\sigma}^\nu-\sigma^\nu\check{\sigma}^\mu & 0\\
0&\check{\sigma}^\mu\sigma^\nu-\check{\sigma}^\nu\sigma^\mu\ea\right),\\
P^\mu&=&\gamma^\mu\frac{1+\gamma^5}{2}=\left(\ba{cc} 0& \sigma^\mu \\ 0
&0\ea\right),\;K^\mu=\gamma^\mu\frac{1-\gamma^5}{2}=\left(\ba{cc} 0& 0
\\ \check{\sigma}^\mu &0\ea\right)\ea\label{confalgamma}\ee
fulfils the commutation relations (\ref{conformalgebra}). These
are the Lie algebra generators of the fundamental representation
of the four cover of $SO(4,2)$:
\be SU(2,2)=\left\{g=\left(\ba{cc} A& B
\\ C &D\ea\right)\in {\rm Mat}_{4\times 4}(\mathbb C):  g^\dag \Gamma g=\Gamma,
\det(g)=1\right\},\label{su22} \ee
with $\Gamma$ a ${4\times 4}$ hermitian form of signature
$(++--)$. In particular, taking $\Gamma=\gamma^5$, the $2\times 2$
complex matrices $A,B,C,D$ in (\ref{su22}) satisfy the following
restrictions:
\be g^{-1}g=I_{4\times 4}\Leftrightarrow \left\{\ba{r} D^\dag
D-B^\dag B=\sigma^0\\ A^\dag A-C^\dag C=\sigma^0\\ A^\dag B-C^\dag
D=0,\ea\right.\label{mim}\ee
together with those of $gg^{-1}=I_{4\times 4}$. In this article we
shall work with $G=U(2,2)$ instead of $SO(4,2)$ and we shall use a
set of complex coordinates to parametrize $G$. This parametrization will be adapted
to the non-compact complex Grassmannian $\mathbb D=G/H$ of the
maximal compact subgroup $H=U(2)^2$. It can be obtained through a
block-orthonormalization process with metric $\Gamma=\gamma^5$ of
the matrix columns of:
\be \left(\ba{cc} \sigma^0& 0
\\ Z^\dag &\sigma^0\ea\right)\rightarrow g=\left(\ba{cc} \sigma^0& Z
\\ Z^\dag &\sigma^0\ea\right)\left(\ba{cc} \Delta_1& 0
\\ 0 &\Delta_2\ea\right), \left\{ \ba{l} \Delta_1=(\sigma^0-ZZ^\dag)^{-1/2}\\
\\ \Delta_2=(\sigma^0-Z^\dag Z)^{-1/2}\ea\right..
\nn\ee
Actually, we can identify
\be Z=Z(g)=BD^{-1}, Z^\dag=Z^\dag(g)=CA^{-1},
\Delta_1=(AA^\dag)^{1/2},\Delta_2=(DD^\dag)^{1/2}\label{zeta}.\ee
From (\ref{mim}), we obtain the positive-matrix conditions
$AA^\dag>0$ and $DD^\dag>0$, which are equivalent to:
\be \sigma^0-ZZ^\dag>0,\; \sigma^0-Z^\dag Z>0.\label{positive}\ee
Moreover, from the top condition of (\ref{mim}), we arrive at the
determinant restriction:
\be \det(ZZ^\dag)=\det(B^\dag B)\det(\sigma^0+B^\dag
B)^{-1}<1,\label{dl1}\ee
which, together with
$\det(\sigma^0-ZZ^\dag)=1-\tr(ZZ^\dag)+\det(ZZ^\dag)>0$, implies
that $\tr(ZZ^\dag)<2$. Thus, we can identify the symmetric complex
Cartan domain
\be \mathbb D=G/H=\{Z\in {\rm Mat}_{2\times 2}(\mathbb C):
\sigma^0-ZZ^\dag>0\} \label{cartandomain}\ee
with an open subset of the eight-dimensional ball with radius
$\sqrt{2}$. Moreover, the compactified Minkowski space $\mathbb M$
is the Shilov boundary $U(2)=\{Z\in {\rm Mat}_{2\times 2}(\mathbb
C): Z^\dag Z=ZZ^\dag=\sigma^0\}$ of $\mathbb D$.

There is a one-to-one mapping from $\mathbb D$ onto the future
tube domain
\be \mathbb T=\{W=X+iY\in {\rm Mat}_{2\times 2}(\mathbb C):\,
Y>0\},\label{tube}\ee
of the complex Minkowski space $\mathbb C^{1,3}$, with
$X=x_\mu\sigma^\mu$ and $Y=y_\mu\sigma^\mu$ hermitian matrices and
$Y>0\Leftrightarrow y^0>\|\vec{y}\|$. This map is given by the
Cayley transformation and its inverse:
\be Z\to W(Z)=i(\sigma^0-Z)(\sigma^0+Z)^{-1},\;\;W\to
Z(W)=(\sigma^0-iW)^{-1}(\sigma^0+iW).\label{Cayley}\ee
This is the 3+1-dimensional analogue of the usual map form the
unit disk onto the upper half-plane in two dimensions. Actually,
the forward tube domain $\mathbb T$ is naturally homeomorphic to
the quotient $G/H$ in a new realization of $G$ in terms of
matrices $f$ which preserve $\Gamma=\gamma^0$, instead of
$\Gamma=\gamma^5$; that is, $f^\dag\gamma^0 f=\gamma^0$. Both
realizations of $G$ are related by the map
\be g\to f=\Upsilon g \Upsilon^{-1}, \;\;\Upsilon=\frac{1}{\sqrt{2}}\left(\ba{cc} \sigma^0& -\sigma^0\\
\sigma^0& \sigma^0 \ea\right).\label{upsilon}\ee
We shall come again to this ``forward tube domain'' realization
later on Section \ref{tubedomainsec}.

Let us proceed by giving a complete local parametrization of $G$ adapted
to the fibration $H\to G\to \mathbb D$. Any element $g\in G$ (in the
present patch, containing the identity element) admits the Iwasawa
decomposition
\be g=\left(\ba{cc} A& B
\\ C &D\ea\right)=\left(\ba{cc} \Delta_1& Z\Delta_2
\\ Z^\dag\Delta_1 &\Delta_2\ea\right)\left(\ba{cc} U_1& 0
\\ 0 &U_2\ea\right),\label{Iwasawa}\ee
where the last factor
\be U_1=\Delta_1^{-1}A, U_2=\Delta_2^{-1}D \nn\ee
belongs to $H$; i.e., $U_1,U_2\in U(2)$. Likewise, a
parametrization of any $U\in U(2)$ (in a patch containing the
identity), adapted to the quotient $\mathbb S^2=U(2)/U(1)^2$, is
(the Hopf fibration)
\be U=\left(\ba{cc} a& b
\\ c &d\ea\right)=\left(\ba{cc} \delta & z\delta
\\ -\bar{z}\delta & \delta\ea\right)\left(\ba{cc} e^{i\alpha}& 0
\\ 0 & e^{i\beta}\ea\right),\label{Iwasawa2}\ee
where $z=b/d\in \overline{\mathbb C}\simeq \mathbb S^2$ (the
one-point compactification of $\mathbb C$ by inverse stereographic
projection), $\delta=(1+z\bar{z})^{-1/2}$ and $e^{i\alpha}=a/|a|,
e^{i\beta}=d/|d|$.

Sometimes it will be more convenient
for us to use the following compact notation for the sixteen
coordinates of $U(2,2)$:
\be \left\{\ba{cccc} \alpha_1 &z_1 & Z_{11} & Z_{12}  \\ -\bar{z}_1&
\beta_1 & Z_{21} & Z_{22} \\ \bar{Z}_{11} &  \bar{Z}_{21} & \alpha_2 & z_2 \\
\bar{Z}_{12} & \bar{Z}_{22} & -\bar{z}_2&
\beta_2\ea\right\}= \left\{\ba{cccc} x^{1}_{1}& x^{1}_{2}& x^{1}_{3} & x^{1}_{4} \\
x^{2}_{1}& x^{2}_{2}& x^{2}_{3} & x^{2}_{4}
\\x^{3}_{1} & x^{3}_{2} & x^{3}_{3} &x^{3}_{4} \\ x^{4}_{1}&x^{4}_{2} &x^{4}_{3} &x^{4}_{4}
\ea\right\}=\{x^{\alpha}_{\beta}(g)\},\label{xz}\ee
The set of coordinates $\{x^{\alpha}_{\beta}\}$ is adapted to the new Lie
algebra basis of step operator matrices
$({X}_{\alpha}^{\beta})_{\mu}^{\nu}\equiv
\delta_{\alpha}^{\nu}\delta^{\beta}_{\mu}$  fulfilling the commutation
relations:
\be \left[{{X}}_{\alpha_1}^{\beta_1},{{X}}_{\alpha_2}^{\beta_2}\right]=
\delta_{\alpha_1}^{\beta_2}{{X}}_{\alpha_2}^{\beta_1} -
\delta_{\alpha_2}^{\beta_1}{{X}}_{\alpha_1}^{\beta_2}, \label{bosreal}\ee
and the usual orthogonality properties:
\be
\tr(X_{\alpha}^{\beta}X_{\gamma}^{\rho})=\delta_{\alpha}^{\rho}\delta_{\gamma}^{\beta}.
\nn\ee
The Cartan (maximal Abelian) subalgebra $u(1)^4\subset {\cal G}$ is made
of diagonal operators $\{ X_{\alpha}^{\alpha}, \alpha=1,\dots,4\}$.

Another realization of the conformal Lie algebra that will be
useful for us is the one given in terms of left- and
right-invariant vector fields, as generators of right- and
left-translations of $G$,
\be
[\mathcal{U}^R_{g}\psi](g')=\psi(g'g),\;\;[\mathcal{U}^L_g\psi](g')=\psi(g^{-1}g'),
\label{rightleftrep}\ee
on complex functions $\psi:G\to\mathbb C$, respectively. Denoting by
\be \theta^L=-ig^{-1}dg=\theta^{\alpha}_{\beta}
X_{\alpha}^{\beta}=\theta^{\alpha\mu}_{\beta\nu}dx^{\nu}_{\mu}X_{\alpha}^{\beta}\label{Maurer}\ee
the left-invariant Maurer-Cartan 1-form, the left-invariant vector fields
$L_{\alpha}^{\beta}$ are defined by duality
$\theta^{\alpha}_{\beta}(L_{\rho}^{\sigma})=\delta^\alpha_\rho\delta_\beta^\sigma$.
The same applies to right-invariant 1-forms $\theta^R=-idgg^{-1}$ in
relation with right-invariant vector fields $R_{\alpha}^{\beta}$. They can
also be computed through the group law $g''=g'g$ as:
\be L_{\alpha}^{\beta}(g)\equiv \left.\frac{\partial
x^{\mu}_{\nu}(gg')}{\partial
x^{\alpha}_{\beta}(g')}\right|_{g'=1}\frac{\partial}{\partial
x^{\nu}_{\mu}(g)},\;\; R_{\alpha}^{\beta}(g)\equiv \left.\frac{\partial
x^{\mu}_{\nu}(g'g)}{\partial
x^{\alpha}_{\beta}(g')}\right|_{g'=1}\frac{\partial}{\partial
x^{\nu}_{\mu}(g)}. \label{vectorfields}\ee
The quadratic Casimir operator (\ref{Casimir}) now adopts the compact
form:
\be
C_2=L_{\alpha}^{\beta}L_{\beta}^{\alpha}=R_{\alpha}^{\beta}R_{\beta}^{\alpha}.\nn\ee
Both sets of vector fields will be essential in our quantization
procedure, the first ones ($L$) as generators of gauge transformations and
the second ones ($R$) as the symmetry operators of our theory.

\section{\label{NLSMsec}Non-linear sigma models on $G$}

The actual Lagrangian for quantum mechanical geodesic free motion on $G$,
as a configuration space, is given by:
\be \cL_G(g,\dot g)=\frac{1}{2}\tr (\vartheta^L)^2=\frac{1}{2}
\vartheta^{\alpha}_{\beta}\vartheta^{\beta}_{\alpha}=\frac{1}{2}{\rm
g}_{\mu\rho}^{\nu\sigma}(x)\dot{x}^{\mu}_{\nu}\dot{x}^{\rho}_{\sigma},\label{LG}\ee
where we are denoting by
\[\vartheta^L=-ig^{-1}\dot
g=\vartheta^{\alpha}_{\beta}X_{\alpha}^{\beta}=\vartheta^{\alpha\nu}_{\beta\mu}\dot{x}^{\mu}_{\nu}X_{\alpha}^{\beta}\]
the restriction of (\ref{Maurer}) to trajectories $g=g(t)$ and writing the
natural metric on $G$, ${\rm
g}_{\mu\rho}^{\nu\sigma}=\vartheta^{\alpha\nu}_{\beta\mu}\vartheta_{\alpha\rho}^{\beta\sigma}$,
in terms of vielbeins $\vartheta^\alpha_\beta$. The equations of motion
derived from (\ref{LG}) are: $\dot{\vartheta}^L=0$, which can be converted
into the standard form of geodesic motion
\[\ddot{x}^a+\Gamma^a_{bc}(x)\dot{x}^b\dot{x}^c=0\]
by introducing the Levi-Civita connection $\Gamma^a_{bc}$ [here we used an
alternative indexation $a=(\alpha\beta)=1,\dots,16$, to simplify
expressions]. The phase space of this theory is the cotangent bundle
$T^*G$, which can be identified with the product of $G$ and its Lie
algebra ${\cal G}$ in a suitable way.

It can be shown that the Lagrangian (\ref{LG}) is $G$-invariant
under both: left- and right-rigid transformations, $g(t)\to
g'g(t)$ and $g(t)\to g(t)g'$, respectively; that is, $\cL_G$ is
chiral. This chirality is partially broken when we reduce the
dynamics from $G$ to certain cosets $G/G^0$, with $G^0$ the
isotropy subgroup of a given Lie algebra element of the form
\be X_0=\sum_{\alpha=1}^4\lambda_\alpha
X_{\alpha}^{\alpha}\label{X0}\ee
(with $\lambda_\alpha$ some real constants) under the adjoint
action $X_0\to g X_0 g^{-1}$ of $G$ on its Lie algebra ${\cal G}$.
Actually, the new Lagrangian on $G/G^0$ can be written as a
``partial trace'':
\be \cL_{G/G^0}(g,\dot g)=\frac{1}{2}\tr_{G/G^0}
(\vartheta^L)^2\equiv
\frac{1}{2}\tr([X_0,\vartheta^L])^2=\frac{1}{2}
\sum_{\alpha,\beta=1}^{N}(\lambda_\alpha-\lambda_\beta)^2\vartheta^{\alpha}_{\beta}\vartheta^{\beta}_{\alpha}.
\label{LGH}\ee
For example, choosing $X_0=\frac{\lambda}{2} \gamma^5=\lambda D$
(the dilation) we have $G^0=H=U(2)^2$ (the maximal compact
subgroup) and $G/G^0$ the eight-dimensional domain $\mathbb D$.
For $\lambda_\alpha\not=\lambda_\beta, \forall
\alpha,\beta=1,\dots,4$, the isotropy subgroup of $X_0$ is the
maximal Abelian subgroup $G^0=U(1)^4$ and $G/G^0=\mathbb F$ is a
twelve-dimensional ``pseudo-flag'' (non-compact) manifold. It is
obvious that $\cL_{G/G^0}$ is still invariant under general rigid
left-transformations $g(t)\to g'g(t)$. However, this Lagrangian is
now singular or, equivalently:
{\prop The Lagrangian (\ref{LGH}) is gauge invariant under local
right-transformations
\be g(t)\to g(t)g_0(t), \;\forall g_0(t)\in G^0\label{gauge1}\ee }
\textbf{Proof:} we have that:
\[ \vartheta^L=-ig^{-1}\dot g\to
\vartheta'^L=-ig_0^{-1}g^{-1}(\dot g g_0+g\dot g_0)=g_0^{-1}\vartheta^L
g_0-ig_0^{-1}\dot g_0\]
and
\[[X_0,\vartheta'^L]=g_0^{-1}[X_0,\vartheta^L]g_0,\]
since $G^0$ is the isotropy subgroup of $X_0$, which means
$[X_0,g_0]=0=[X_0,\dot g_0]$. The cyclic property of the trace completes
the proof $\blacksquare$

We have considered so far $G/G^0$ as a configuration space. In
this article, we shall be rather interested in $G/G^0$ as a phase
space. For example, we shall consider $\mathbb D$ [or the tube
domain (\ref{tube}) of the complex Minkowski space $\mathbb
C^{1,3}$] as a (complex) phase space of four-position $x^\mu$ and
four-momenta $y^\nu$, in itself. This situation will require a new
singular Lagrangian of the form:
\be \cL(g,\dot
g)=\tr(X_0\vartheta^L)=\sum_{\alpha=1}^{4}\lambda_\alpha\vartheta^{\alpha}_{\alpha}.\label{L}\ee
Again, this Lagrangian is left-$G$-invariant under rigid transformations.
The difference now is that it is linear in velocities $\dot x$. Moreover,
we shall prove that:
{\prop The Lagrangian (\ref{L}) is gauge (semi-)invariant under local
right- transformations
\be g(t)\to g(t)g_0(t), \;\forall g_0(t)\in G^0\label{gauge2}\ee
up to a total time derivative, i.e.,
\be \cL\to \cL+\Delta \cL,\;\;\Delta \cL=-i\tr(X_0g_0^{-1}\dot
g_0)=\frac{d\tau}{dt}, \; \tau=\sum_{\alpha=1}^4 \lambda_\alpha
x^{\alpha}_{\alpha}.\label{seminvariante}\ee }
\textbf{Proof:} We shall just consider the two important cases for us:
\begin{enumerate}
\item $\lambda_\alpha\not=\lambda_\beta, \forall
\alpha\not=\beta\Rightarrow G^0=U(1)^4, G/G^0=\mathbb F$ \item
$X_0=\lambda D=\frac{\lambda}{2}\gamma^5\Rightarrow G^0=H=U(2)^2,
G/G^0=\mathbb D$.
\end{enumerate}
For the first case, any $g_0\in G^0$ can be written as
$g_0=\exp(ix^{\alpha}_{\alpha}X_{\alpha}^{\alpha})$ and $\dot g_0=ig_0\dot
x^{\alpha}_{\alpha}X_{\alpha}^{\alpha}$ because $G^0$ is Abelian;
therefore
\be \Delta \cL=-i\tr(X_0g_0^{-1}\dot g_0)=\sum_{\beta=1}^4
\lambda_\beta \dot x^{\alpha}_{\alpha}\tr(X_{\beta}^{\beta}
X_{\alpha}^{\alpha})=\sum_{\alpha=1}^4\lambda_\alpha \dot
x^{\alpha}_{\alpha}.\nn\ee
For the second case, $g_0=\exp(i\varphi I+i\tau'
D+i\omega^{\mu\nu}M_{\mu\nu})=\left(\ba{cc} U_1& 0\\ 0
&U_2\ea\right)\in H$. Disregarding the trivial global phase
$\varphi$, it is clear that for dilations $g_0=d_0=e^{i\tau' D}$ we
have $\dot d_0=i\dot\tau' De^{i\tau' D}$ and
\be-i\tr(X_0d_0^{-1}\dot d_0)=\lambda\dot\tau
\tr(D^2)=\lambda\dot\tau'=\dot\tau,\nn\ee
where $\tau\equiv\lambda\tau'$. For Lorentz transformations
$g_0=m_0=\exp(i\omega^{\mu\nu}M_{\mu\nu})$ we have $\Delta \cL=0$
since $\tr(DM_{\mu\nu})=0$, which is a direct consequence of the
orthogonality properties of the Pauli matrices
$\tr(\sigma^\mu\sigma^\nu)=2\delta^{\mu\nu}$. $\blacksquare$

{\rem \label{dilhamil} We can always fix the gauge to $\tau(t)=t$.
In the case $X_0=\lambda D$, this implies that the dilation
operator $D$ will play the role of the Hamiltonian of the quantum
theory. The replacement of time translations by dilations as
dynamical equations of motion has been considered in
\cite{conformecontract} and in \cite{dilatatiempo} when quantizing
field theories on space-like Lorentz-invariant hypersurfaces
$x^2=x^\mu x_\mu=\tau^2=$constant. In other words, if one wishes
to proceed from one surface at $x^2=\tau_1^2$ to another at
$x^2=\tau_2^2$, this is done by scale transformations; that is,
$D$ is the evolution operator in a proper time $\tau$ $\square$.}

\section{\label{quantizationsec}Quantum mechanics in the phase space $G/G^0$}

We shall see that the constants $\lambda_\alpha$ label the (lowest
weight) irreducible representations of $G$ on which the Hilbert
space of our theory is constructed. There are several ways of
seing that the values of $\lambda_\alpha$ are quantized. One way
is through the path integral method. To examine this explicitly,
consider the transition amplitude from an initial point $g_1$ at
$t=t_1$ to a final point $g_2$ at $t=t_2$. For each path $g(t)$
connecting $g_1$ and $g_2$, there are many gauge equivalent paths
\be g'(t)=g(t)g_0(t), \; g_0(t)\in G^0, \;
g_0(t_1)=g_0(t_2)=1\nn\ee
that must contribute to the sum of the path integral with the same
amplitude, that is:
\be e^{i\int_{t_1}^{t_2} dt \cL(g,\dot g)}=e^{i\int_{t_1}^{t_2} dt
\cL(g',\dot g')}=e^{i\int_{t_1}^{t_2} dt \cL(g,\dot
g)}e^{i\int_{t_1}^{t_2} dt \Delta \cL(g,\dot g)}\Rightarrow
e^{i\int_{t_1}^{t_2} dt \Delta \cL(g,\dot g)}=1.\nn\ee
Using (\ref{seminvariante}), the last expression can be written as
$\exp(i(\tau(t_2)-\tau(t_1))=1$ which, together with the fact that
\[g_0(t_{1,2})=e^{i\sum_\alpha x^{\alpha}_{\alpha}(t_{1,2})}=1 \Leftrightarrow x^{\alpha}_{\alpha}(t_{1,2})=2\pi n^{\alpha}_{1,2},\,\,
n_{1,2}^{\alpha}\in\mathbb Z,\]
means that $\lambda_\alpha$ must be an
integer number. Considering coverings of $G$, one can relax the integer to a half-integer condition, as
happens with $SU(2)$ in relation with $SO(3)$.

Other alternative way to the path-integral description of
realizing the integrality of $\lambda_\alpha$ is though the following
operator (representation-theoretic) description. At the quantum level, finite-right gauge transformations like
(\ref{gauge2}) induce constraints on ``physical'' wave functions $\psi(g)$
as:
\be \psi(gg_0)=\mathcal{U}^{\lambda}_0({g_0})\psi(g),\;\;g_0\in
G^0\label{constraintfinite}\ee
where we are allowing $\psi$ to transform non-trivially according
to a representation $\mathcal{U}^{\lambda}_0$ of $G^0$ of index
$\lambda$. This could be seen as a generalization of the original
Dirac approach to the quantization of constrained systems (where
$\mathcal{U}^{\lambda}_0$ is taken to be trivial) which allows new
inequivalent quantizations labelled by $\lambda_\alpha$ (see e.g.
\cite{Ramirez,McMullan,FracHall,Landsmann} for several approaches
to the subject). The finite constraint condition
(\ref{constraintfinite}) can be written in infinitesimal form as
\be
 L^\alpha_\alpha\psi
 =\lambda_\alpha\psi, \;
 \alpha=1,\dots,4,\label{constraint1}
 \ee
where we have used the fact that left-invariant vector fields
(\ref{vectorfields}) are generators of finite right-transformations. In
the parametrization $\{x^\alpha_\beta\}$, the left-invariant vector fields
$L_\alpha^\beta$ fulfill the same commutation relations as the step
operator matrices (\ref{bosreal}). Therefore, when acting on
physical/constrained states (\ref{constraint1}), they satisfy creation and
annihilation harmonic-oscillator-like commutation relations:
\be
[L_\alpha^\beta,L_\beta^\alpha]=(\lambda_\beta-\lambda_\alpha)\;\;\;
(\mathrm{no\, sum\, on}\,\alpha,\beta). \nn\ee
We shall work in a holomorphic picture, which means that constrained wave
functions (\ref{constraint1}) will be further restricted by holomorphicity conditions:
\be
 L_\alpha^\beta\psi=0, \ \forall \alpha>\beta=1,2,3.\label{constraint2}
\ee
In fact, looking at (\ref{vectorfields}), for $g\in G$ near the identity we
have $L_\alpha^\beta(g)\sim
\partial/\partial x^\alpha_\beta$ so that $L_\alpha^\beta\psi=0$
means, roughly speaking, that $\psi(g)$ does not depend on the
variables $x^\alpha_\beta, \alpha>\beta=1,2,3$ in (\ref{xz}), that
is, $\psi$ is holomorphic. The complementary option
$L_\alpha^\beta\psi=0, \forall \beta>\alpha=1,2,3$ then leads to
anti-holomorphic functions. Those readers familiar with Geometric
Quantization \cite{Souriau,Woodhouse} will identify the constraint
equations (\ref{constraint1}) and (\ref{constraint2}) as
\emph{polarization} conditions (see also \cite{GAQ} for a Group
Approach to Quantization scheme and \cite{higherpol} for the
extension of first-order polarizations to higher-order
polarizations), intended to reduce the left-representation
$\mathcal{U}^L$ (\ref{rightleftrep}) of $G$, on complex wave
functions $\psi$, to $G/G^0$. Also, the constraints
(\ref{constraint1}) and (\ref{constraint2}) are exactly the
defining relations of a lowest-weight representation.

\subsection{\label{scalarsec}Conformal scalar quantum particles}
Firstly we shall consider the (spin-less) case
$\lambda_1=\lambda_2=-\lambda_3=-\lambda_4\equiv -\lambda/2$, that
is, $X_0=\frac{\lambda}{2}\gamma^5=\lambda D$, and we shall call
$\lambda$ the \emph{conformal, scale or mass dimension}. In this
case the gauge group is the maximal compact subgroup
$G^0=H=U(2)^2$ and the phase space is the eight-dimensional domain
$\mathbb D=G/G^0$.

\subsubsection{Constraint conditions and physical wave functions}
The constraint conditions
(\ref{constraint1}) can now be enlarged to
\be
D^L\psi=-\frac{1}{2}(L^1_1+L^2_2-L^3_3-L^4_4)\psi=\lambda\psi,\;
M^L_{\mu\nu}\psi=0.\label{constrained3}\ee
which renders translation ($P_\mu$) and acceleration ($K_\nu$) generators
into conjugated variables. In fact, the last commutator of
(\ref{conformalgebra}), on constrained (physical) wave functions
(\ref{constrained3}), gives:
\be \lb K_\mu^L,P_\nu^L\rb\psi = 2\lambda\eta_{\mu\nu}\psi, \ee
which states that $K_\mu$ and $P_\mu$ can not be simultaneously
measured, the conformal dimension $\lambda$ playing here the role
of the Planck constant $\hbar$. Note that $K_\mu$ and $P_\mu$ are
conjugated but not \emph{canonically} conjugated as such. We
address the reader to Refs. \cite{confposition1,confposition2} for
other definitions of quantum observables associated with positions
in space-time, namely \be
X_\mu=M_{\nu\mu}\cdot\frac{P^\nu}{P^2}+D\cdot
\frac{P_\mu}{P^2}\label{positionop}\ee (dot means symmetrization),
fulfilling canonical commutation relations
$[X_\mu,P_\nu]=\eta_{\mu\nu}$ inside the conformal (enveloping)
algebra (\ref{conformalgebra}).

A further restriction
\be K_\mu^L\psi=0\label{constrained4}\ee
selects the holomorphic (``position'') representation. Indeed, let
us prove that:
{\thm \label{solupolath}The general solution to
(\ref{constrained3}) and (\ref{constrained4}) can be factorized
as:
\be \psi_\lambda(g)={\cal W}_\lambda(g)\phi(Z),\label{solupola}\ee
where the ``ground state''
\bea {\cal W}_\lambda(g)&=&\det(D)^{-\lambda}=\det(\sigma^0-Z^\dag
Z)^{\lambda/2}\det(U_2)^{-\lambda}\nn\\
&=&(1-\tr(Z^\dag Z)+\det(Z^\dag
Z))^{\lambda/2}\det(U_2)^{-\lambda}\label{vac}\eea
is a particular solution of
(\ref{constrained3},\ref{constrained4}) and $\phi$ is the general
solution for the trivial representation $\lambda=0$ of $G^0=H$
(actually, an arbitrary, analytic holomorphic function of $Z$),
for the decomposition (\ref{Iwasawa}) of an element $g\in G$.}

\ni\textbf{Proof:} A generic proof (also valid for other symmetry
groups) that the general solution of
(\ref{constrained3},\ref{constrained4}) admits a factorization of
the form (\ref{solupola}) can be found in the Proposition 3.3 of
\cite{acha}. Here we shall just prove that (\ref{solupola}) is a
solution of (\ref{constrained3},\ref{constrained4}). Indeed, by
applying a finite right translation (\ref{rightleftrep}) on ${\cal
W}_\lambda(g)$:
\bea [{\cal U}^R_{g'}{\cal W}_\lambda](g)&=&{\cal W}_\lambda(g
g{'})=\det(D'')^{-\lambda}=\det(CB{'}+DD{'})^{-\lambda}\nn\\
&=&\det(D')^{-\lambda}\det(CZ'+D)^{-\lambda},
\label{transvacright}\eea
we see that ${\cal W}_\lambda(gg')$ is not affected by
translations by $Z'^\dag=Z^\dag(g')=C'A'^{-1}$. Infinitesimally,
it means that $K^L_\mu {\cal W}_\lambda(g)=0$, according to the
lower-triangular choice of the generator $K_\mu$ in
(\ref{confalgamma}). For Lorentz transformations we have $B'=0=C'$
and $\det(A')=1=\det(D')$ and therefore ${\cal
W}_\lambda(gg')={\cal W}_\lambda(g)$, that is $M^L_{\mu\nu}{\cal
W}_\lambda(g)=0$. For dilations we have $B'=0=C'$ and
$A'=e^{i\tau/2}\sigma^0=D'^\dag$, which gives ${\cal
W}_\lambda(gg')=e^{i\lambda\tau}{\cal W}_\lambda(g)$ or $D^L{\cal
W}_\lambda(g)=\lambda {\cal W}_\lambda(g)$ for small $\tau$. It
remains to prove that $\phi(Z)$ is the general solution of
(\ref{constrained3},\ref{constrained4}) for $\lambda=0$. From
(\ref{zeta}) we have
\be Z''=Z(gg')=B''D''^{-1}=(AB'+BD')(CB'+DD')^{-1},\label{rightz}\ee
which is not affected by $C'$ and gives $Z''=Z$ for dilations and
Lorentz transformations ($B'=0$) $\blacksquare$

{\rem \label{BRP-CPT}In the last theorem, we are implicitly
restricting ourselves to gauge transformations $g'\in S(U(2)^2)$,
which means $\det(g')=\det(U_1U_2)=1$. If we allow for
transformations $g'\in U(2)^2$ with $\det(g')\not=1$ (like
$e^{i\alpha}I$) and we want them to leave physical wave functions
strictly invariant $\psi(gg')=\psi(g)$ (i.e., we restrict
ourselves to representations with
$\lambda_1+\lambda_2+\lambda_3+\lambda_4=0$), we must choose a
symmetrical form for the ground state
\be  {\cal W}_\lambda(g)=\det(A^\dag)^{-\lambda/2}
\det(D)^{-\lambda/2}=\det(\sigma^0-Z^\dag
Z)^{{\lambda}/{2}}\det(U_1^\dag)^{-\lambda/2}\det(U_2)^{-\lambda/2},\label{vacsym}\ee
which reduces to (\ref{vac}) for $\det(U_1U_2)=1$.

Moreover, instead of (\ref{constrained4}), we could have chosen the
complementary constraint $P^L_\mu\psi=0$ which would have led us
to a anti-holomorphic (``acceleration'') representation
$\psi_\lambda(g)=\check {\cal W}_\lambda(g)\phi(Z^\dag)$ with the
new ground state
\be  \check{\cal W}_\lambda(g)=\det(A)^{-\lambda/2}
\det(D^\dag)^{-\lambda/2}=\det(\sigma^0-Z^\dag
Z)^{{\lambda}/{2}}\det(U_1)^{-\lambda/2}\det(U_2^\dag)^{-\lambda/2},\label{vacsym2}\ee
which, for $g\in SU(2,2)$, reduces to:
\be\check {\cal
W}_\lambda(g)=\det(A)^{-\lambda}=\det(\sigma^0-ZZ^\dag)^{\lambda/2}\det(U_1)^{-\lambda}=\overline{{\cal
W}_\lambda(g)}.\nn\ee
Therefore, the BRP-like symmetry  $K^L_\mu\leftrightarrow P^L_\mu,
D^L\to -D^L$ in (\ref{BRPconf}) manifest here as a charge
conjugation and time reversal (CT) operations. See later on
Section \ref{comments} for more details on a ``BRP-CPT
connection'' proposal inside the conformal group. $\square$}

\subsubsection{Irreducible representation, Haar measure and Bergman kernel}
The finite left-action of $G$ on physical wave functions
(\ref{solupola}),
\bea [{\cal
U}^L_{g'}\psi_\lambda](g)&=&\psi_\lambda(g'^{-1}g)=\det(D(g'^{-1}g))^{-\lambda}\phi(Z')\nn\\
&=& {\cal W}_\lambda(g)\det(D'^\dag-B'^\dag Z)^{-\lambda}\phi(Z'),\label{repre}\\
Z'&\equiv&Z(g'^{-1}g)=(A'^\dag Z-C'^\dag)(D'^\dag-B'^\dag Z)^{-1}, \nn\eea
provides a unitary irreducible representation of $G$ under the invariant
scalar product
\be\la \psi_\lambda|\psi'_\lambda\ra=\int_G
d\mu^L(g)\overline{\psi_\lambda(g)}\psi'_\lambda(g)\label{scalarprod}\ee
given trough the left-invariant Haar measure [the exterior product
of left-invariant one-forms (\ref{Maurer})] which can be decomposed as:
\be\ba{rcl}
d\mu^L(g)&=&c\bigwedge_{\alpha,\beta=1}^4\vartheta^\alpha_\beta=
c\det(\vartheta^{\alpha\mu}_{\beta\nu})\bigwedge_{\mu,\nu=1}^4dx^\nu_\mu\\
&=&c\left.d\mu(g)^L\right|_{G/H}\left.d\mu^L(g)\right|_{H},\\
\left.d\mu^L(g)\right|_{G/H}&=&
\det(\sigma^0-ZZ^\dag)^{-4}|dZ|,\\
\left.d\mu(g)\right|_{H}&=& dv(U_1)dv(U_2),\ea\label{Haarmeasure}\ee
where we are denoting by $dv(U)$ the Haar measure on $U(2)$,
which can be in turn decomposed as:
\bea
dv(U)&\equiv& \left.dv(U)\right|_{U(2)/U(1)^2} \left.dv(U)\right|_{U(1)^2},\nn\\
\left.dv(U)\right|_{U(2)/U(1)^2}
 &=& \left.dv(U)\right|_{\mathbb{S}^2}\equiv ds(U)=(1+z\bar
z)^{-2}|dz|,\label{haarmeasures2}\\
\left.dv(U)\right|_{U(1)^2}&\equiv& d\alpha d\beta.\nn \eea
We have used the Iwasawa decomposition of an element $g$ given in
(\ref{Iwasawa},\ref{Iwasawa2}) and denoted by $|dz|$ and $|dZ|$
the Lebesgue measures in  $\mathbb C$ and $\mathbb C^4$,
respectively. The normalization constant
\be
c=\pi^{-4}(\lambda-1)(\lambda-2)^2(\lambda-3)\left(\frac{(2\pi)^3}{2}\right)^{-2}\label{normconst}\ee
is fixed so that the ground state (\ref{vac}) is normalized, i.e.
$\scprod{{\cal W}_\lambda}{{\cal W}_\lambda}=1$ (see Appendix B of Ref.
\cite{EMSMTA} for orthogonality properties), the factor
${(2\pi)^3}/{2}$ actually being  the volume $v(U(2))$. The
scalar product (\ref{scalarprod}) is finite as long as
$\lambda\geq 4$.

The infinitesimal generators of (\ref{repre}) are
the right-invariant vector fields $R^\alpha_\beta(g)$ in
(\ref{vectorfields}) and constitute the operators (observables) of
our quantum theory. For example, from the general expression
(\ref{repre}), we can compute the finite left-action of dilations $g'=e^{i\tau D}$
($B'=0=C'$ and $A'=e^{-i\tau/2}\sigma^0=D'^\dag$) on physical wave
functions,
\be \psi_\lambda(g'g)=e^{i\lambda \tau}{\cal
W}_\lambda(g)\phi(e^{i\tau}Z),\nn\ee
or infinitesimally:
\bea
D^R\psi_\lambda(g)&=&-\frac{1}{2}(R^1_1+R^2_2-R^3_3-R^4_4)\psi_\lambda(g)=
{\cal
W}_\lambda(g)\left(\lambda+\sum_{i,j=1}^2Z_{ij}\frac{\partial}{\partial
Z_{ij}}\right)\phi(Z)\nn\\ &\equiv& {\cal W}_\lambda(g) D_\lambda
\phi(Z),\label{dilaction}\eea
where we have defined the restriction of the dilation operator on holomorphic functions as:
\be
D_\lambda\equiv\lambda+\sum_{i,j=1}^2Z_{ij}\frac{\partial}{\partial
Z_{ij}},\label{dilactionrest}\ee
for future use. As we justified in Remark \ref{dilhamil}, the dilation generator
$D^R$ plays the role of the Hamiltonian operator of this theory
\be \hat{\cal H}=-i\frac{\partial}{\partial \tau}=
D^R.\label{hamiltonianscalar}\ee
The conformal or mass dimension $\lambda$ can be then interpreted
as the zero point (vacuum) energy and the corresponding
eigenfunctions are homogeneous polynomials $\phi_n(Z)$ of a
certain degree (eigenvalue) $n$, according to Euler's theorem. We
shall come back to this question later in Theorem \ref{GMSMT}.

Let us introduce bracket notation and write:
\be {\cal W}_{\lambda}(g)\equiv\langle g|\lambda,0\rangle=\langle
\lambda,0|{\cal
U}^L_{g^{-1}}|\lambda,0\rangle,\;\psi_\lambda(g)\equiv\langle
g|\psi_\lambda\rangle. \ee
Here we are implicitly making use of the Coherent-States machinery
(see e.g. \cite{Perelomov,Klauder}). Actually, we are denoting by
$|g\rangle\equiv {\cal U}^L_{g}|\lambda,0\rangle$ the set of
vectors in the orbit of the ground (``fiducial'') state
$|\lambda,0\rangle$ (the lowest-weight vector) under the left
action of the group $G$ (this set is called a family of
\emph{covariant coherent states} in the literature
\cite{Perelomov,Klauder}). We can easily calculate the coherent
state overlap:
\bea \scprod{g'}{g}&=&\bra{\lambda,0}{\cal
U}^L_{g'^{-1}g}|\lambda,0\rangle={\cal
W}_\lambda(g^{-1}g')=\det(D(g^{-1}g'))^{-\lambda}=\det(D^\dag
D'-B^\dag B')^{-\lambda}\nn\\
&=&\det(D^\dag)^{-\lambda}\det(\sigma^0-(BD^{-1})^\dag
B'D'^{-1})^{-\lambda} \det(D')^{-\lambda}\nn\\ &=& \overline{{\cal
W}_\lambda(g)}\det(\sigma^0-Z^\dag Z')^{-\lambda}{\cal
W}_\lambda(g').\label{cohov}\eea
The set of coherent states $\{\ket{g}, g\in G\}$ constitutes a tight frame (see \cite{EMSMTA} for a proof in
the context of Conformal Wavelets) with resolution of unity:
\be 1=\int_G d\mu(g) \ket{g}\bra{g}.\nn\ee
Actually, the coherent state overlap (\ref{cohov}) is a reproducing kernel
satisfying the integral equation of a projector operator
\be
\scprod{g}{g''}=\int_Gd\mu^L(g')\scprod{g}{g'}\scprod{g'}{g''}\nn\ee
and the propagator equation
\be
\psi_\lambda(g')=\int_Gd\mu^L(g)\scprod{g'}{g}\psi_\lambda(g).\nn\ee
Since the ground state ${\cal W}_\lambda$ is a fixed common factor of all
the wave functions (\ref{solupola}), we could factor it out and define the
restricted left-action
\be[{\cal U}^\lambda_{g'}\phi](Z)\equiv {\cal
W}_\lambda^{-1}(g)[{\cal
U}^L_{g'}\psi_\lambda](g)=\det(D'^\dag-B'^\dag
Z)^{-\lambda}\phi(Z')\equiv  \phi'(Z)\label{reprerest}\ee
of $G$ on the arbitrary (holomorphic) part $\phi$ of
$\psi_\lambda$, instead of (\ref{repre}). In standard (induced)
representation theory, the factor $\det(D'^\dag-B'^\dag
Z)^{-\lambda}$ is called a ``multiplier'' (Radon-Nicodym
derivative) and fulfils cocycle properties. For the representation
(\ref{reprerest}) of $G$ on holomorphic functions $\phi(Z)$ to be
unitary, the left-$G$-invariant Haar measure (\ref{Haarmeasure})
has to be accordingly modified as:
\be d\mu_\lambda(Z,Z^\dag)\equiv c_\lambda|{\cal
W}_\lambda(g)|^2\left.d\mu^L(g)\right|_{G/H}=c_\lambda
\det(\sigma^0-ZZ^\dag)^{\lambda-4} |dZ|,\label{projintmeasure}\ee
where $\left.d\mu^L(g)\right|_{G/H}$ in (\ref{Haarmeasure})
is the projection of the
left-$G$-invariant Haar measure $d\mu^L(g)$  onto $G/H$.
Roughly speaking, we are integrating out the coordinates of $H$ and redefining
the normalization constant $c$ in (\ref{normconst}) as
$c_\lambda=c/v(U(2))=\pi^{-4}(\lambda-1)(\lambda-2)^2(\lambda-3)$
so that the unit constant function $\phi(Z)=1$ (the ground state)
is normalized (see \cite{EMSMTA} for orthogonality properties). As
before, we could also introduce a modified bracket notation
$\phi(Z)\equiv \scprodp{Z}{\phi}$ and a new set $\{\ketp{Z},
Z\in\mathbb D\}$ of coherent states in the Hilbert space ${\cal
H}_\lambda({\mbD})=L^2(\mathbb D,d\mu_\lambda)$ of analytic square-integrable holomorphic functions
$\phi$ on $\mathbb D$. The new coherent state overlap
$\scprodp{Z}{Z'}$ is nothing but the so called reproducing
Bergman's kernel $K_\lambda(Z,Z')$. It is related to (\ref{cohov})
by:
\be K_\lambda(Z',Z)=\scprodp{Z'}{Z}=\frac{\scprod{g'}{g}}{ {\cal
W}_\lambda(g')\overline{{\cal W}_\lambda(g)}}=\det(\sigma^0-Z^\dag
Z')^{-\lambda}.\label{renormkernel}\ee
We notice that, unlike $\ket{g}$, the coherent state $\ketp{Z}$ is
not normalized. In fact,
\be{\cal K}_\lambda(Z,Z^\dag)\equiv\ln
\scprodp{Z}{Z}\label{Kahlerpot}\ee
is nothing but the K\"ahler potential, which
defines $\mathbb D$ as a K\"ahler manifold with local complex
coordinates $Z=z_\mu\sigma^\mu$, an Hermitian Riemannian metric ${\rm g}$ and
a corresponding closed two-form $\omega$
\be ds^2={\rm g}^{\mu\nu}dz_{\mu} \odot d\bar z_{\nu},\;\;
\omega=-i{\rm g}^{\mu\nu}dz_{\mu}\wedge d\bar z_{\nu}, \;\; {\rm
g}^{\mu\nu}\equiv\frac{\partial^2{\cal K}_\lambda}{\partial
z_{\mu}\partial \bar z_{\nu}},\label{Kahlerform}\ee
where $\odot$ denotes symmetrization. We shall come back to the Riemannian structure of $\mbD$ and $\mathbb T$
and the connection with the BRP later on Section \ref{tubedomainsec}.

\subsubsection{Schwinger's theorem, orthonormal basis and closure relations}
As already commented after Eq.
(\ref{dilaction}), we are interested in calculating an orthonormal
basis of ${\cal H}_\lambda(\mathbb D)$ made of Hamiltonian
eigenfunctions $\varphi_J(Z)\equiv \scprodp{Z}{\lambda, J}$, where
$J$ denotes a set of indices. This orthonormal basis would provide
us with a new resolution of the identity
\be 1=\sum_J \ketp{\lambda,J}\brap{\lambda,J}.\nn\ee
Actually, we shall identify $\varphi_J(Z)$ by looking at the expansion of the
Bergman's kernel
\be K_\lambda(Z',Z)=\scprodp{Z'}{Z}=\sum_J \scprodp{Z'}{\lambda,
J}\scprodp{\lambda, J}{Z}=\sum_J
\varphi_J(Z')\overline{\varphi_J(Z)}.\nn\ee
Thus, the Bergman's kernel plays here the role of a generating function. To be more precise:

{\thm\label{GMSMT}  The infinite set of polynomials
\be
\varphi_{q_1,q_2}^{j,m}(Z)=\sqrt{\frac{2j+1}{\lambda-1}\binom{m+\lambda-2}{\lambda-2}\binom{m+2j+\lambda
-1}{\lambda-2}}\det(Z)^{m}\cD^{j}_{q_1,q_2}(Z),\label{basisfunc}\ee
with
\bea
\cD^{j}_{q_1,q_2}(Z)=\sqrt{\frac{(j+q_1)!(j-q_1)!}{(j+q_2)!(j-q_2)!}}
\sum_{p=\max(0,q_1+q_2)}^{\min(j+q_1,j+q_2)}\binom{j+q_2}{p}\binom{j-q_2}{p-q_1-q_2}\nn\\
\times z_{11}^p
z_{12}^{j+q_1-p}z_{21}^{j+q_2-p}z_{22}^{p-q_1-q_2}\label{Wignerf}\eea
the standard Wigner's $\cD$-matrices ($j$ is a non-negative half-integer), verifies the following
closure relation (the reproducing Bergman kernel):
\be\sum_{j\in\mbN/2}\sum^{\infty}_{m=0}\sum^{j}_{q_1,q_2=-j}
\overline{\varphi_{q_1,q_2}^{j,m}({Z})}\varphi_{q_1,q_2}^{j,m}(Z')=\frac{1}{\det(\sigma^0-Z^\dag
Z')^\lambda}\label{closure}\ee
and constitute an orthonormal basis of ${\cal H}_\lambda(\mathbb
D)$.}

This theorem has been proven in \cite{EMSMTA}. It turns out to be rooted in a extension of the Schwinger's formula:

{\thm\label{MSMT} {\rm (Schwinger's Master Theorem)} The identity
 \be
 \sum_{j\in\mbN/2}t^{2j}\sum^{j}_{q=-j}\cD^{j}_{qq}(X)=\frac{1}{\det(\sigma^0-tX)}\label{MSMTF}\ee
holds for any $2\times 2$ matrix $X$, with $t$ an arbitrary parameter.}

The abovementioned extension of the Theorem
\ref{MSMT} can be stated as:

{\thm \label{EMSMT} {\rm ($\lambda$-Extended Schwinger's Master
Theorem)} For every $ \lambda  \in \mathbb{N}, \lambda \geq 2$ and
every $2\times 2$ complex matrix $X$ the
following identity holds:
\bea
 \sum_{j\in\mbN/2}\frac{2j+1}{\lambda-1}\sum^{\infty}_{m=0}t^{2j+2m}
\binom{m+\lambda-2}{\lambda-2}\binom{m+2j+\lambda -1}{\lambda-2}
 \det(X)^{m}\sum^{j}_{q=-j}\cD^{j}_{qq}(X)\nn\\ ={\det(\sigma^0-tX)^{-\lambda }}.\label{EMSMTF}\eea
}
We address the interested reader to Ref. \cite{EMSMTA} for a complete proof.


\ni\textbf{Scketch of proof of Theorem \ref{GMSMT}:} Assuming the validity of (\ref{EMSMTF}) and replacing $tX=Z^\dag Z'$ in it, we have:
\bea \sum_{j\in\mbN/2}\frac{2j+1}{\lambda-1}\sum^{\infty}_{m=0}
\binom{m+\lambda-2}{\lambda-2}\binom{m+2j+\lambda -1}{\lambda-2}
 \det(Z^\dag Z')^{m}\sum^{j}_{q=-j}\cD^{j}_{qq}(Z^\dag Z')\nn\\ =
 \frac{1}{\det(\sigma^0-Z^\dag Z')^{\lambda }}\label{lemma2}\,.\eea
Using determinant and Wigner's $\cD$-matrix rules
\be
 \det(Z^\dag Z')^{n}\sum^{j}_{q=-j}\cD^{j}_{qq}(Z^\dag Z')=
 \det(Z^\dag)^{n}\det(Z')^{n}\sum^{j}_{q_2,q_1=-j}\overline{\cD^{j}_{q_1q_2}(Z)}\cD^{j}_{q_1q_2}(Z')\nn\ee
and the definition of the functions (\ref{basisfunc}), we see that (\ref{lemma2})
reproduces (\ref{closure}). On the other hand, the number of linearly independent polynomials
$\prod_{i,j=1}^2 z_{ij}^{n_{ij}}$ of fixed degree of homogeneity
$n=\sum_{i,j=1}^2n_{ij}$ is $(n+1)(n+2)(n+3)/6$, which coincides
with the number of linearly independent polynomials
(\ref{basisfunc}) with degree of homogeneity $n=2m+2j$. This proves
that the set of polynomials (\ref{basisfunc}) is a basis for
analytic functions $\phi\in {\cal H}_\lambda(\mathbb D_4)$. Moreover, this basis turns
out to be orthonormal under the
projected integration measure (\ref{projintmeasure}). We address the interested reader
to the Appendix B of Ref. \cite{EMSMTA} for a proof.$\blacksquare$

{\rem The set (\ref{basisfunc}) constitutes a basis of Hamiltonian
eigenfunctions with energy eigenvalues $E_{n}^\lambda$ (the
homogeneity degree) given by:
\be
\hat{\cal H}_\lambda\varphi_{q_1,q_2}^{j,m}=E_{n}^\lambda\varphi_{q_1,q_2}^{j,m},\;\;
E_{n}^\lambda=\lambda+n,\;\; n=2j+2m,\label{energyspectrum}\ee
with $\hat{\cal H}_\lambda=D_\lambda$ defined in
(\ref{dilactionrest}). Each energy level $E_{n}^\lambda$ is then
$(n+1)(n+2)(n+3)/6$ times degenerated. The spectrum is equi-spaced
and bounded from below, with $E_{0}^\lambda=\lambda$ playing the
role of a zero-point energy. At this stage it is interesting to
compare our Hamiltonian choice with others in the literature like
\cite{Mulak} studying a $SU(2,2)$-harmonic oscillator on the phase
space $\mathbb D$. In this case the quantum Hamiltonian is chosen
to be the Toeplitz operator corresponding to the square of the
distance with respect to the $SU(2,2)$-invariant K\"ahler metric
(\ref{Kahlerform}) on the phase space $\mathbb D$. $\square$

}

\subsection{\label{spinningsec}Conformal spinning  quantum particles}

Let us use the following notation for
\be X_0=\sum_{\alpha=1}^4\lambda_\alpha
X_{\alpha}^{\alpha}=\lambda D+s_1 \varSigma^{(3)}_{1}+s_2
\varSigma^{(3)}_{2}+ \kappa I,\label{X0s}\ee
where
\[\varSigma^{(3)}_{1}=X_1^1-X^2_2=\left(\ba{cc} \sigma^3& 0\\ 0 &0\ea\right),\;\;
\varSigma^{(3)}_{2}=X_3^3-X^4_4=\left(\ba{cc} 0& 0\\ 0 &
\sigma^3\ea\right)\]
stand for the third spin components and $I$ the $4\times 4$
identity matrix. The identification (\ref{X0s}) implies that the
spin labels of the representation of the subgroup $SU(2)^2$ are
$s_1\equiv(\lambda_1-\lambda_2)/2$ and
$s_2\equiv(\lambda_3-\lambda_4)/2$. The conformal dimension is
$\lambda=(\lambda_3+\lambda_4-\lambda_1-\lambda_2)/2$ and
$\kappa=(\lambda_1+\lambda_2+\lambda_3+\lambda_4)/4$ is the
(trace) $U(1)$ quantum number. We shall choose, without lost of
generality, $\kappa=0$, which means that $\lambda$ remains integer
(as in the spin-less case) and that we are restricting ourselves
to representations of $SU(2,2)\subset U(2,2)$.

{\thm The general solution to (\ref{constraint1}) and
(\ref{constraint2}) can be factorized as:
\be \psi_\lambda^{s_1,s_2}(g)={\cal
W}_\lambda^{s_1,s_2}(g)\phi(Z,z_1,z_2),\label{solupola2}\ee
where the ground state
\be\ba{rcl} {\cal W}_\lambda^{s_1,s_2}(g)&=&\det(A^\dag)^{-\lambda_s/2}
\det(D)^{-\lambda_s/2}\cD^{s_1}_{s_1,s_1}(U_1^\dag)\cD^{s_2}_{-s_2,-s_2}(U_2)\\
&=& \det(A^\dag)^{-\lambda_s/2}\det(D)^{-\lambda_s/2}\bar{a}_1^{2s_1}d_2^{2s_2} \\ &=&
\det(\Delta_1U_1^\dag)^{-\lambda_s/2}\det(\Delta_2U_2)^{-\lambda_s/2}(\delta_1e^{-i\alpha_1})^{2s_1}(\delta_2e^{i\beta_2})^{2s_2}\\
&=&\det(\sigma^0-Z^\dag Z)^{\frac{\lambda_s}{2}}(1+\bar
z_1z_1)^{-s_1}(1+\bar
z_2z_2)^{-s_2}\\ & & \times e^{-i\alpha_1(2s_1-\lambda_s/2)}e^{i\beta_1\lambda_s/2}e^{-i\alpha_2\lambda_s/2}
e^{i\beta_2(2s_2-\lambda_s/2)},\ea\label{vac2}\ee
with $\lambda_s\equiv\lambda-s_1-s_2$, is a particular solution of (\ref{constraint1},\ref{constraint2})
and $\phi$ is the general solution for the trivial representation
$\lambda_\alpha=0$ of $G^0=U(1)^4$ (actually, an arbitrary,
analytic holomorphic function of $Z,z_1,z_2$), for the
decomposition (\ref{Iwasawa},\ref{Iwasawa2}) of an element $g\in
G$.}

\ni\textbf{Proof:} On the one hand, from (\ref{transvacright}) we
conclude that the factors $\det(D)^{-\lambda}$ and
$\cD^{s_2}_{-s_2,-s_2}(U_2)$, with $U_2=(DD^\dag)^{-1/2}D$ fulfill
the holomorphicity conditions (\ref{constraint2}) for
$(\beta,\alpha)=(1,3)$, $(2,3)$, $(1,4)$, $(2,4)$. Moreover,
$U_1^\dag=A^\dag(AA^\dag)^{-1/2}$ and we have that
\be A''^\dag=A(gg')^\dag=A'^\dag A^\dag+C'^\dag
B^\dag=A'^\dag(A^\dag+(C'A'^{-1})^\dag B^\dag)=A'^\dag(A^\dag+Z'
B^\dag)\nn\ee
is not affected by $Z'^\dag=C'A'^{-1}$ either, according to the
definition (\ref{zeta}). On the other hand, for $g'\in H$ we have
that
\bea \bar a''&=&\bar a(gg')=\bar a\bar a'+\bar b\bar c'=\bar a'(\bar a-z'\bar b)\nn\\
 d''&=&d(gg')=cb'+dd'=d'(d+z'c)\nn\eea
are not affected by $\bar z'=- c'/a'=\bar b'/\bar d'$, according
to the definition (\ref{Iwasawa2}). This proves that the ground
state (\ref{vac2}) fulfills the holomorphicity conditions
(\ref{constraint2}) for $(\beta,\alpha)=(1,2)$, $(3,4)$. It
remains to prove the gauge conditions (\ref{constraint1}) or their
finite counterpart (\ref{constraintfinite}) for $g_0\in
G^0=U(1)^4$. Finite right
(gauge) dilations $g_0=e^{i\tau D}$ leave ${\cal
W}_\lambda^{s_1,s_2}(gg_0)=e^{i\lambda\tau}{\cal
W}_\lambda^{s_1,s_2}(g)$ invariant up to the phase
$\cU^\lambda_0(g_0)=e^{i\lambda\tau}$ (a character of $G^0$),
where we have used that $\det(\cdot)$ and $\cD^s(\cdot)$ are
homogeneous of degree 2 and $2s$, respectively. Infinitesimally,
it means that
$D^L\psi_\lambda^{s_1,s_2}=\lambda\psi_\lambda^{s_1,s_2}$.  For $g_0^{(1,2)}=e^{i\alpha\varSigma^3_{1,2}}$
the ground state transforms as
expected:
\[{\cal
W}_\lambda^{s_1,s_2}(gg_{0}^{(1,2)})=\cU^\lambda_0(g_{0}^{(1,2)}){\cal
W}_\lambda^{s_1,s_2}(g),\;\;\cU^\lambda_0(g_{0}^{(1,2)})=e^{2is_{1,2}\alpha}
.\]
Infinitesimally, it means that
\be\varSigma^{L(3)}_{1,2}{\cal W}_\lambda^{s_1,s_2}=2s_{1,2}{\cal
W}_\lambda^{s_1,s_2},\;
\left\{\ba{l}\varSigma^{L(3)}_{1}\equiv L^1_1-L^2_2,\\
\varSigma^{L(3)}_{2}\equiv
L^3_3-L^4_4.\ea\right.\label{sigmaL3}\ee
Moreover, one can easily check that ${\cal
W}_\lambda^{s_1,s_2}(gg_0)={\cal
W}_\lambda^{s_1,s_2}(g)$ for diagonal $U(1)$ transformations $g_0=e^{i\theta}I$, that is, $\kappa=0$. Finally, using
similar arguments to those employed in (\ref{rightz}), we can assert that
$z'_{1,2}=z_{1,2}(gg_0)=z_{1,2},\,\forall g_0\in G^0$, which ends up proving the gauge
conditions (\ref{constraintfinite})$\blacksquare$

{\rem Instead of (\ref{constraint2}), we could have chosen the
complementary constraint $ L_\alpha^\beta\psi=0, \ \forall
\alpha<\beta$  which would have led us to a anti-holomorphic
representation. $\square$}

As in Eq. (\ref{repre}), we can compute the finite left-action of
$G$ on physical wave functions (\ref{solupola2}). In particular,
for the case of dilations $g'=e^{i\tau' D}$ (i.e., $B'=0=C'$ and
$A'=e^{-i\tau'/2}\sigma^0=D'^\dag$) we have:
\be \psi_\lambda^{s_1,s_2}(g'^{-1}g)=e^{i\lambda \tau'}{\cal
W}_\lambda^{s_1,s_2}(g)\phi(e^{i\tau'}Z,z_1,z_2),\nn\ee
or infinitesimally:
\be D^R\psi_\lambda^{s_1,s_2}(g)= {\cal
W}_\lambda^{s_1,s_2}(g)\left(\lambda+\sum_{i,j=1}^2Z_{ij}\frac{\partial}{\partial
Z_{ij}}\right)\phi(Z,z_1,z_2).
\label{dilaction2}\ee
Comparing this expression with (\ref{dilaction}), we realize that
the spin coordinates $z_1,z_2$ do not contribute to the degree of
homogeneity of $\phi$ under dilations, as they correspond
to ``internal'' (versus space-time-momentum) degrees of freedom.

As in the previous subsection, we can introduce a modified bracket
notation $\phi(Z,z_1,z_2)\equiv \scprodp{Z,z_1,z_2}{\phi}$ and a
set $\{\ketp{Z,z_1,z_2}, Z\in\mathbb D, z_1,z_2\in\mathbb C\}$ of
coherent states in the Hilbert space ${\cal
H}_\lambda^{s_1,s_2}(\mathbb F)$ of analytic measurable
holomorphic functions $\phi$ on the twelve-dimensional pseudo-flag
manifold $\mathbb F=U(2,2)/U(1)^4$, locally $\mathbb
D\times\overline{\mathbb C}^2$, with integration measure
\be d\mu_\lambda^{s_1,s_2}(Z,z_1,z_2;Z^\dag,\bar z_1,\bar
z_2)\equiv d\mu_{\lambda_s}(Z,Z^\dag)\frac{2s_1+1}{\pi}ds(U_1)
\frac{2s_2+1}{\pi}ds(U_2),\ee
where
$d\mu_{\lambda_s}(Z,Z^\dag)$ and $ds(U)$ are defined in
(\ref{projintmeasure}) and (\ref{haarmeasures2}), respectively. Note
that the square-integrability condition $\lambda\geq 4$ in ${\cal
H}_\lambda(\mathbb D)$ becomes $\lambda_s\geq 4$ in ${\cal
H}_\lambda^{s_1,s_2}(\mathbb F)$. The constant factor
${(2s_1+1)}/{\pi}$ is introduced so that the following set of
functions is normalized.

  {\thm\label{GMSMT3}  The infinite set of polynomials
\bea
\check\varphi_{j,q_1,q_2}^{m,m_1,m_2}(Z,z_1,z_2)&\equiv&(-1)^{m_1+s_1}
\varphi_{q_1,q_2}^{j,m}(Z)
\frac{\cD^{s_1}_{s_1,-m_1}(U_1^\dag)\cD^{s_2}_{m_2,-s_2}(U_2)}{\cD^{s_1}_{s_1,s_1}(U_1^\dag)\cD^{s_2}_{-s_2,-s_2}(U_2)}
\nn\\
&=&\varphi_{q_1,q_2}^{j,m}(Z)\sqrt{\binom{2s_1}{m_1+s_1}\binom{2s_2}{m_2+s_2}}z_1^{m_1+s_1}z_2^{m_2+s_2},
\label{basisfunc2}\eea
(with $\varphi_{q_1,q_2}^{j,m}$ in (\ref{basisfunc}) replacing
$\lambda\to\lambda_s$) provides an orthonormal basis of ${\cal
H}_\lambda^{s_1,s_2}(\mathbb F)$. The closure relation:
\bea\sum^{\infty}_{j\in\mbN/2}\sum^{\infty}_{m=0}\sum^{j}_{q_1,q_2=-j}\sum^{s_1}_{m_1=-s_1}\sum^{s_2}_{m_2=-s_2}
\check\varphi_{j,q_1,q_2}^{m,m_1,m_2}(Z',z_1',z_2')\overline{\check\varphi_{j,q_1,q_2}^{m,m_1,m_2}({Z,z_1,z_2})}\nn\\
= \scprodp{Z',z_1',z_2'}{Z,z_1,z_2}\label{closure2}\eea
gives the reproducing Bergman's kernel for spinning particles:
\bea
K_\lambda^{s_1,s_2}(Z',z'_1,z'_2;Z,z_1,z_2)&=&\scprodp{Z',z_1',z_2'}{Z,z_1,z_2}
\nn\\
&=&\det(\sigma^0-Z^\dag Z')^{-\lambda_s}(1+\bar
z_1z_1')^{2s_1}(1+\bar z_2z_2')^{2s_2}.\eea
 }

\ni \textbf{Proof:} Assuming the orthonormality of
(\ref{basisfunc}) (see Appendix B of Ref. \cite{EMSMTA}), and
realizing that
\[\int_{\mathbb C}\sqrt{\binom{2s}{m}}\bar
z^{m}\sqrt{\binom{2s}{m'}}z^{m'}\frac{2s+1}{\pi}ds(U)=\delta_{m,m'},\;
m,m'=0,\dots,2s,\]
we prove the orthonormality of the functions (\ref{basisfunc2}).
Moreover, the number of linearly independent polynomials
$\prod_{i,j=1}^2 z_{ij}^{n_{ij}}\prod_{i=1}^2 z_{i}^{n_{i}}$ with
$0\leq n_i\leq 2s_i$ and fixed $n=\sum_{i,j=1}^2n_{ij}$ is
$(2s_1+1)(2s_2+1)(n+1)(n+2)(n+3)/6$, which coincides with the
number of linearly independent polynomials (\ref{basisfunc2}) with
degree of homogeneity $n=2m+2j$ in the coordinates $Z$. This proves that the set
of polynomials (\ref{basisfunc2}) is a basis for analytic
functions ${\cal H}_\lambda^{s_1,s_2}(\mathbb F)$.

It just remains to prove the closure relation (\ref{closure2}).
This proof reduces to that of Theorem \ref{GMSMT} when noting the
binomial identity $\sum^{2s}_{m=0}\tbinom{2s}{m}(\bar
zz')^m=(1+\bar zz')^{2s}$ or the Wigner $\cal D$-matrix property
\[\sum^{s}_{n=-s}
\cD^{s}_{sn}(U)\cD^{s}_{ns}(U')=\cD^{s}_{ss}(UU').\blacksquare\].

{\rem At this point it is interesting to compare our construction
with others in the literature like \cite{Ruhl1}, where the
proposed basis functions
\be
\Phi_{j,q_1,q_2}^{m,m_1,m_2}(A,D,Z)=\cD^{j_1}_{j_1,m_1}(A^T)\cD^{j_2}_{m_2,j_2}(D)
\varphi_{q_1,q_2}^{j,m}(Z) \ee
do not form an orthogonal set unless a coupling between orbital
angular momentum $j$ with spin $j_1,j_2$ by means of
Clebsch-Gordan coefficients is made:
\be
\tilde\Phi_{j,j_1,j_2}^{m,p_1,p_2}=\sum_{m_1,m_2,q_1,q_2}C(j,q_1;s_1,m_1-s_1|j_1,p_1)
C(j,q_2;s_2,m_2-s_2|j_2,p_2)\Phi_{j,q_1,q_2}^{m,m_1,m_2}. \nn\ee
Moreover, the fact that $U_1=(AA^\dag)^{-1/2}A$ and
$U_2=(DD^\dag)^{-1/2}D$ introduces a new contribution of
$\cD^{j_1}(A)$ and $\cD^{j_2}(D)$ to the integration measure
$d\mu_\lambda^{j_1,j_2}$, with respect to $\cD^{j_1}(U_1)$ and
$\cD^{j_2}(U_2)$, such that the square-integrability condition
becomes $\lambda\geq 4+2j_1+2j_2$.$\square$
 }

The Hamiltonian of our spinning particle is $\hat{\cal
H}=-i\frac{\partial}{\partial \tau}$ with $\tau$ given by
(\ref{seminvariante}). Its expression in terms of right-invariant
vector fields $R^\alpha_\beta$ is then
\be\hat{\cal
H}=\sum_{\alpha=1}^4\frac{1}{\lambda_\alpha}R^\alpha_\alpha=
\rho_0D^R+\rho_1\varSigma^{R(3)}_1+\rho_2\varSigma^{R(3)}_2+\rho_3
I,\ee
with:
\be
\rho_0=\frac{4\lambda(\lambda^2-s_1^2-s_2^2)}{(\lambda^2-4s_1^2)(\lambda^2-4s_2^2)},\;\rho_1=\frac{-4s_1}{\lambda^2-4s_1^2},\;
\rho_2=\frac{-4s_2}{\lambda^2-4s_2^2},\;
\rho_3=\frac{4\lambda(s_2^2-s_1^2)}{(\lambda^2-4s_1^2)(\lambda^2-4s_2^2)},\nn\ee
and $\varSigma^{R(3)}_{1,2}$ the right-invariant version of
(\ref{sigmaL3}). In order to compare with the spin-less case, we
can always renormalize
\be\hat{\cal H}\to \hat{\cal H}/\rho_0=\hat{\cal
H}'=D^R+\varrho_1\varSigma^{R(3)}_1+\varrho_2\varSigma^{R(3)}_2+\varrho_3
I,\label{hamilspin}\ee
with $\varrho_\alpha=\rho_\alpha/\rho_0$.  We can interpret
$\varrho_{1,2}$  as constant ``magnetic fields'' (oriented along
the ``$z$'' direction) coupled to the spin degrees of freedom
$\varSigma^{R(3)}_{1,2}$. The set (\ref{basisfunc2}) constitutes a
basis of eigenfunctions of the Hamiltonian (\ref{hamilspin}) with
eigenvalues (energy levels)  given by:
\be E_{n,q_1,q_2}^{\lambda,
m_1,m_2}=\lambda+\varrho_3+n+\varrho_1(m_1+q_1)+\varrho_2(m_2+q_2),\;\;\;
n=2j+2m.\ee
Comparing this energy eigenvalues with the energy spectrum
(\ref{energyspectrum}) of the spin-less Hamiltonian $\hat{\cal
H}=D^R$, we realize that the zero-point energy has been shifted
from $\lambda$ to $\lambda+\varrho_3-s_1\varrho_1-s_2\varrho_2$.
Like in the (anomalous) \textit{Zeeman effect}, the introduction
of spin leads to an splitting of a spin-less spectral line
$E^\lambda_n$ into $(2s_1+1)(2s_2+1)$ components in the presence
of a ``static magnetic field'' $\varrho_{1,2}$.

\section{Relation with the tube domain realization}\label{tubedomainsec}

In this section we shall translate some expressions obtained from the
complex Cartan domain (\ref{cartandomain}) into the forward tube domain
(\ref{tube}), where we enjoy more (Minkowskian) intuition. We shall
restrict ourselves to the scalar case, since it is representative of the
more general case.

\subsection{Tube domain as a homogeneous space of $SU(2,2)$}

As we have already said, the forward tube domain $\mathbb T$ is
naturally homeomorphic to the quotient $G/H$ in the realization of
$G$ in terms of matrices
\be f=\left(\ba{cc} R& iS
\\ -iT &Q\ea\right)\ee
which preserve $\Gamma=\gamma^0$, instead of $\Gamma=\gamma^5$;
that is, $f^\dag\gamma^0 f=\gamma^0$. Both realizations of $G$ are
related by the map (\ref{upsilon}), which can be explicitly
written as
\be g=\left(\ba{cc} A& B
\\ C &D\ea\right)=\Upsilon^{-1}f\Upsilon=\frac{1}{2}\left(\ba{cc} R+iS-iT+Q &-R+iS+iT+Q\\
-R-iS-iT+Q & R-iS+iT+Q\ea\right).\label{upsilon3} \ee
%
%
The identification of ${\mathbb T}$ with the quotient $G/H$ is
given through
\be W(f)=i(R-iS)(Q+iT)^{-1}.\label{tubecoord}\ee
Hence, the left translation $f'\to ff'$ of $G$ on itself induces a
left action of $G$ on ${\mathbb T}$ given by:
\be W=W(f')\to W'=W(ff')=(RW+S)(TW+Q)^{-1}.\label{tubeaction}\ee
Setting  $W=x_\mu\sigma^\mu$, and making use of the standard
homomorphism (spinor map) between $SL(2,\mbC)$ and $SO^+(3,1)$ given by:
$W'=RWR^\dag \leftrightarrow x'^\mu=\Lambda^{\mu}_\nu x^\nu, R\in SL(2,\mbC), \Lambda^{\mu}_\nu\in SO^+(3,1)$,
the transformations (\ref{confact}) can be recovered from (\ref{tubeaction}) as follows:
\begin{itemize}
 \item[i)] Standard Lorentz transformations, $x'^\mu=\Lambda^{\mu}_\nu(\omega) x^\nu$, correspond to $T=S=0$ and
$R=Q^{-1\dag}\in SL(2,\mbC)$. \item[ii)] Dilations correspond to
$T=S=0$ and $R=Q^{-1}=\rho^{1/2}I $ \item[iii)] Spacetime
translations equal $R=Q=\sigma^0$ and $S=b_\mu\sigma^\mu, T=0$.
\item[iv)] Special conformal transformations correspond to
$R=Q=\sigma^0$ and $T=a_\mu\sigma^\mu, S=0$ by noting that
$\det(\sigma^0 +TW)=1+2ax+a^2x^2$.
\end{itemize}

\subsection{Irreducible representations, Haar measure and Bergman kernel}

Let us see the expression of the wave functions (\ref{solupola})
in the tube domain $\mathbb T$. Performing the change of variables
(\ref{upsilon3}) in (\ref{solupola}) we get
 \be \psi_\lambda(f)= \det(Q+iT)^{-\lambda} 2^{2\lambda}
\det(\sigma^0-iW)^{-\lambda}\phi(Z(W))\equiv
\Omega_\lambda(f)\tilde\phi(W),\label{solupola2-1}\ee
where we have defined a new ground state $\Omega_\lambda$ and a
new function $\tilde\phi$ as:
 \be
\Omega_\lambda(f)\equiv\det(Q+iT)^{-\lambda},\;\tilde\phi(W)\equiv
2^{2\lambda}\det(\sigma^0-iW)^{-\lambda}\phi(Z(W)).\label{isomap}\ee
In the same manner, the coherent-state overlap (\ref{cohov}) can
be cast as
 \be \scprod{f'}{f}=\det(Q^\dag-iT^\dag)^{-\lambda}\det(\frac{i}{2}(W^\dag-W'))^{-\lambda}
 \det(Q'+iT')^{-\lambda}\label{cohov2}\ee

Since the ground state ${\Omega_\lambda}$ is a fixed common factor
of all the wave functions (\ref{solupola2-1}), we can factor it
out (as we did in (\ref{reprerest}) with ${\cal W}_\lambda$) and
define the restricted action
\bea [\tilde{\cal U}^\lambda_{f'}\tilde\phi](Z)&\equiv&
{\Omega}_\lambda^{-1}(f)[{\cal
U}^L_{f'}\psi_\lambda](f)\nn\\&=&\det(R'^\dag-T'^\dag
W)^{-\lambda} \tilde\phi((Q'^\dag W-S'^\dag)(R'^\dag-T'^\dag
W)^{-1})\equiv \tilde\phi'(W)\label{reprerest2}\eea
of $G$ on the arbitrary (holomorphic) part $\tilde\phi$ of
$\psi_\lambda$.  The Radon-Nicodym derivative is now
$\det(R'^\dag-T'^\dag W)^{-\lambda}$. The representation
(\ref{reprerest2}) of $G$ on holomorphic functions $\tilde\phi(W)$
is unitary with respect to the re-scaled integration measure
\be
d\tilde\mu_\lambda(W,W^\dag)\equiv|{\Omega}_\lambda(f)|^2\left.d\tilde\mu^L(f)\right|_{G/H}=\frac{c_\lambda}{2^4}
\det(\frac{i}{2}(W^\dag-W))^{\lambda-4}|dW|,\label{projintmeasure2}\ee
where we are using $|dW|$ as a shorthand for the Lebesgue measure
$\bigwedge_{i,j=1}^2d\Re w_{ij}d\Im w_{ij}$ on $\mathbb T$. To
arrive at (\ref{projintmeasure2}), firstly, we have  performed the
Cayley transformation (\ref{Cayley}) in the projected integration
measure:
\bea \left.d\mu^L(g)\right|_{G/H}&=&c_\lambda
\det(\sigma^0-ZZ^\dag)^{-4}|dZ|\rightarrow\nn\\
\left.d\tilde\mu^L(f)\right|_{G/H}&=&\frac{c_\lambda}{2^4}
\det(\frac{i}{2}(W^\dag-W))^{-4}|dW|,\eea
taking into account that
$\det(\sigma^0-ZZ^\dag)=\det(2i(W^\dag-W))|\det(\sigma^0-iW)|^{-2}$ and
the Jacobian determinant $|dZ|/|dW|=2^{12}|\det(\sigma^0-iW)|^{-8}$, and
secondly, we have written
\be
|{\Omega}_\lambda(f)|^2=\det(Q^\dag-iT^\dag)^{-\lambda}\det(Q+iT)^{-\lambda}=\det(\frac{i}{2}(W^\dag-W))^{\lambda}
\nn\ee
by making use of (\ref{tubecoord}) and its hermitian conjugate.

As in (\ref{renormkernel}), we could also introduce a modified
bracket notation $\tilde\phi(W)\equiv (W|\tilde\phi)$ and a new
set $\{\ketp{W}, W\in\mathbb T\}$ of coherent states in the
Hilbert space $\cal H_\lambda(\mathbb T)$ of analytic measurable
holomorphic functions $\varphi$ on $\mathbb T$. The new coherent
state overlap $\scprodp{W}{W'}$ is the new Bergman's kernel
$\tilde{K}_\lambda(W',W)$. It is related to (\ref{cohov2}) by:
\be \tilde{K}_\lambda(W',W)=\scprodp{W'}{W}=\frac{\scprod{f'}{f}}{
\Omega_\lambda(f')\overline
\Omega_\lambda(f)}=\det(\frac{i}{2}(W^\dag-W'))^{-\lambda}.\label{BergmanKT}\ee
We again notice that, unlike $\ket{f}$, the coherent state
$\ketp{W}$ is not normalized. Now, the K\"ahler potential is $\ln
\scprodp{W}{W}$, which defines $\mathbb T$ as a K\"ahler manifold
too.

The identification (\ref{isomap}) actually provides an isometry between the spaces
of analytic holomorphic functions ${\cal H}_\lambda(\mbD)$ and ${\cal H}_\lambda(\mbT)$. Let us formally state it.

{\prop\label{isometryDC} The correspondence
\be\begin{array}{cccc} \cS_\lambda: & {\cal H}_\lambda(\mbD)
&\longrightarrow&
{\cal H}_\lambda(\mbT)\\
 & \phi & \longmapsto & \cS_\lambda\phi\equiv
\tilde\phi,\end{array} \nn\ee
with
\be
\tilde\phi(W)=2^{2\lambda}\det(I-iW)^{-\lambda}\phi(Z(W))\label{isomap2}\ee
and $Z(W)$ given by the Cayley transformation(\ref{Cayley}), is
an isometry, that is:
\be \la\phi|\phi'\ra_{{\cal H}_\lambda(\mbD)}=
\la\cS_\lambda\phi|\cS_\lambda\phi'\ra_{{\cal
H}_\lambda(\mbT)}.\label{isoec}\ee
Moreover, $\cS_\lambda$ is an intertwiner (equivariant map) of the
representations (\ref{repre}) and (\ref{reprerest2}), that is:
\be \cU_\lambda=\cS_\lambda^{-1}\tilde\cU_\lambda\cS_\lambda.\label{intertwiner}\ee
}

\ni \textbf{Proof:} The isometry property is proven by construction from (\ref{isomap}).
The intertwining relation (\ref{intertwiner}) can be explicitly
written as:
\be\ba{rll} [\cU_\lambda\phi](Z) &=&\det(D^\dag-B^\dag
Z)^{-\lambda}\phi\left((A^\dag Z-C^\dag)(D^\dag-B^\dag
Z)^{-1}\right)=\\
\left[\cS_\lambda^{-1}\tilde\cU_\lambda{\tilde\phi}\right](Z) &=&
\det(I-iW)^\lambda \det(R^\dag-T^\dag
W)^{-\lambda}\det(I-iW')^{-\lambda}\phi(Z(W')),
\ea\label{intertwiner2}\ee
where $W'=(Q^\dag W-S^\dag)(R^\dag-T^\dag W)^{-1}$. On the one
hand, we have that the argument of $\phi$ is:
\bea Z(W')&=&(I+iW')(I-iW')^{-1}\nn\\ &=& \left( (R^\dag-T^\dag
W)+i(Q^\dag W-S^\dag)\right)\left( (R^\dag-T^\dag W)-i(Q^\dag
W-S^\dag)\right)^{-1}\nn\\
&=&\left((R^\dag-iS^\dag )+i(Q^\dag
+iT^\dag)W\right)\left((R^\dag+iS^\dag )-i(Q^\dag
-iT^\dag)W\right)^{-1}.\nn\eea
Taking now into account the map (\ref{upsilon3})  we have:
\bea Z(W') &=&\left((A^\dag-C^\dag )+i(A^\dag
+C^\dag)W\right)\left((D^\dag-B^\dag )-i(D^\dag
+B^\dag)W\right)^{-1}\nn\\
&=&\left(A^\dag(I+iW)-C^\dag(I-iW)\right)\left(D^\dag(I-iW)-B^\dag(I+iW)\right)^{-1}\nn\\
&=&\left(A^\dag Z-C^\dag\right)\left(D^\dag-B^\dag
Z\right)^{-1},\nn\eea
as desired. On the other hand, we have that
\bea(I-iW')(R^\dag-T^\dag W)=(R^\dag-T^\dag W)-i(Q^\dag
W-S^\dag)=(R^\dag+iS^\dag )-i(Q^\dag -iT^\dag)W
 \nn\\  =(D^\dag-B^\dag )-i(D^\dag
 +B^\dag)W
 = D^\dag(I-iW)-B^\dag(I+iW)=(D^\dag-B^\dag
Z)(I-iW)\nn\eea
which implies
\be \det(I-iW)^\lambda \det(R^\dag-T^\dag
W)^{-\lambda}\det(I-iW')^{-\lambda}=\det(D^\dag-B^\dag
Z)^{-\lambda}\nn\ee
That is, the equality of multipliers in (\ref{intertwiner2}). $\blacksquare$

As a direct consequence of Proposition \ref{isometryDC}, the set
of functions defined by
\be \tilde\varphi_{q_1,q_2}^{j,m}(W)\equiv
2^{2\lambda}\det(I-iW)^{-\lambda}\varphi_{q_1,q_2}^{j,m}(Z(W)),\ee
with $\varphi_{q_1,q_2}^{j,m}$ defined in (\ref{basisfunc}),
constitutes an orthonormal basis of
${\cal H}_\lambda(\mbT)$ and the closure relation
\be
\sum_{j\in\mbN/2}\sum^{\infty}_{m=0}\sum_{q,q'=-j}^j\overline{\tilde\varphi_{q',q}^{j,m}(W)}\tilde\varphi_{q',q}^{j,n}(W')
=\det(\frac{i}{2}(W^\dag-W'))^{-\lambda},\label{reprodkernelW}\ee
renders again the reproducing Bergman kernel (\ref{BergmanKT}).

\subsection{K\"ahler structures, Born's reciprocity and maximal
acceleration}

 As we said for the Cartan domain $\mathbb D$ in
(\ref{Kahlerpot}) and (\ref{Kahlerform}), the K\"ahler potential
\be\tilde{\cal K}_\lambda(W,W^\dag)\equiv\ln \scprodp{W}{W}=-\ln
|{\Omega}_\lambda(f)|^2=-\lambda\ln (\Im(w))^2=-\lambda\ln
y^2\label{Kahlerpotubo}\ee
defines $\mathbb T$ as a K\"ahler manifold with local complex
coordinates $W=w_\mu\sigma^\mu, \,w_\mu=x_\mu+iy_\mu$, an
Hermitian Riemannian metric
\be  {\rm g}^{\mu\nu}\equiv\frac{\partial^2\tilde{\cal
K}_\lambda}{\partial w_{\mu}\partial \bar
w_{\nu}}=-\frac{\lambda}{2y^2}\left(\eta^{\mu\nu}-2\frac{y_\mu
y_\nu}{y^2}\right).\label{Kahlerformtubo}\ee
and a corresponding closed two-form $\omega$
\be \omega=-i{\rm g}^{\mu\nu}dw_{\mu}\wedge d\bar
w_{\nu}.\label{Kahlertwoform}\ee
The line element
\be ds^2={\rm g}^{\mu\nu}dw_{\mu}  d\bar
w_{\nu}=-\frac{\lambda}{2y^2}\left(\eta^{\mu\nu}-2\frac{y^\mu
y^\nu}{y^2}\right)(dx_\mu dx_\nu+dy_\mu dy_\nu)\label{lineconf}\ee
turns out to be positive and provides a conformal counterpart of
the Born's line element (\ref{Bornline}). The two-form
(\ref{Kahlertwoform}) defines the Poisson bracket:
\be \left\{a,b\right\}\equiv i{\rm g}_{\mu\nu}\left(\frac{\partial
a}{\partial w_\mu}\frac{\partial b}{\partial \bar
w_\nu}-\frac{\partial b}{\partial w_\mu}\frac{\partial a}{\partial
\bar w_\nu}\right)\ee
for the inverse metric
\be  {\rm
g}_{\mu\nu}=-\frac{2}{\lambda}\left(\eta^{\mu\nu}y^2-2{y_\mu
y_\nu}\right).\label{inversemetric}\ee
so that ${\rm g}_{\mu\nu} {\rm g}^{\nu\rho}=\delta_\mu^\rho$. In
particular, we have that:
\be \left\{x_\mu,y_\nu\right\}= -\frac{1}{2}{\rm
g}_{\mu\nu},\nn\ee
which differs from $\{x_\mu,y_\nu\}=\eta_{\mu\nu}$; that is,
$x_\mu$ and  $y_\nu$ are not ``canonical'' coordinates. However,
we can define a proper conjugate four-momentum $p_\mu\equiv
\lambda y_\mu/y^2$ which gives the desired (canonical) Poisson
bracket
\be \left\{x_\mu,p_\nu\right\}= \eta_{\mu\nu},\ee
as can be checked by direct computation. The line element
(\ref{lineconf}) then becomes:
\be ds^2=-\frac{1}{2\lambda}\left(\eta^{\mu\nu}p^2-2p^\mu
p^\nu\right)(dx_\mu dx_\nu+\frac{\lambda^2}{p^4}dp_\mu
dp_\nu).\label{lineconf2}\ee
Note the close resemblance between the coordinates,
$dx^\mu(K^\nu)=-2x^\mu x^\nu +x^2\eta^{\mu\nu}$, of the vector
field $K^\nu$ in (\ref{confvf}) and the metric coefficients
$(-2p^\mu p^\nu+p^2\eta^{\mu\nu})$ in (\ref{lineconf2}) under the
interchange $x_\mu\leftrightarrow p_\mu$. The line element
(\ref{lineconf2}) of the (curved) manifold $\mathbb T$ is the
conformal counterpart of the  Born's line element
(\ref{Bornline}) in the (flat) complex Minkowski space $\mathbb
C^{1,3}$, both of them considered as phase spaces of relativistic (conformal)
particles. Concerning the extension of BRP to the case
of curved spacetimes, see also \cite{Castro3} for the construction a reciprocal general
relativity theory as a local gauge theory of the quaplectic group of
\cite{Low1,Low2}.

Remember that one could deduce the existence of a maximal
acceleration from the positivity of the Born's line element
(\ref{amaxBorn}). The existence of a maximal acceleration inside
the conformal group does not seem to be apparent from
(\ref{lineconf2}), although there are other arguments supporting
the existence of a bound $a_{\rm max}$ for proper accelerations. One of them was given time
ago in Ref. \cite{conforme-maxac}, where the authors analyzed the
physical interpretation of the singularities, $1+2a x+a^2 x^2=0$,
of the conformal transformations to a uniformly accelerating frame
[last transformation in (\ref{confact})]. When applying the
transformation to an extended object of size $\ell$, an
upper-limit to the proper acceleration, $a_{\rm max}\simeq
c^2/\ell$, is shown to be necessary in order that the tenets of
special relativity not be violated (see \cite{conforme-maxac} for
more details).

In a coming paper \cite{Unruhconf}, we shall provide an
alternative proof of the existence of a maximal acceleration
inside the conformal group. It is related to the Unruh effect
(vacuum radiation in uniformly accelerated frames) and turns out
to be a consequence of the finiteness of the radiated energy
(black body spectrum). Contrary to other approaches to the Unruh
effect, a bound for the proper acceleration does not necessarily
imply a bound for the temperature.

\section{Comments and outlook\label{comments}}

We have revised the use of complex Minkowski 8-dimensional space
(more precisely, the domains $\mathbb D$ and $\mathbb T$) as a
base for the construction of conformal-invariant quantum (field)
theory, either as a phase space or a configuration space [the last
case related to Lagrangians of type (\ref{LGH})]. We have followed
a gauge-invariant Lagrangian approach (of nonlinear sigma-model
type) and we have used a generalized Dirac method for the
quantization of constrained systems, which resembles in some
aspects the particular approach to quantizing coadjoint orbits of
a group $G$ developed in, for instance, \cite{Bal}.

One could think of these 8-dimensional domains as the replacement
of space-time at short distances or high momentum transfers, as it
is implicit in the original BRP \cite{Born1,Born2}, the standard
relativity theory being then the limit $\ell_{\rm min}\to 0$.
Group-theoretical revisions of the BRP, replacing the Poincaré by
the Canonical (or Quaplectic) group of reciprocal relativity, have
been proposed in \cite{Low1,Low2}. In this article we put a
(conformal) BRP-like forward, as a natural symmetry inside the
conformal group $SO(4,2)$ and the replacement of space-time by the
8-dimensional conformal domain $\mathbb D$ or $\mathbb T$ at short
distances. Actually, we feel tempted to establish a connection
between \textit{holomorphicity$\leftrightarrow$chirality} and
BRP$\leftrightarrow$CPT symmetry inside the conformal group.
Indeed, the definition of $P_\mu$ and $K_\mu$ in
(\ref{confalgamma}) is linked to the right- and left-handed
projectors $(1+\gamma^5)/2$ and $(1-\gamma^5)/2$, respectively.
According to the (conformal) BRP-like symmetry (\ref{BRPconf}),
conformal physics is symmetric under the interchange
$P_\mu\leftrightarrow K_\mu$, as long as we perform a proper-time
reversal $D\to -D$. On the other hand, $P_\mu\leftrightarrow
K_\mu$ entails a swapping of chirality
$(1+\gamma^5)/2\leftrightarrow(1-\gamma^5)/2$, a complex
conjugation $\psi_\lambda(g)\leftrightarrow
\check\psi_\lambda(g)=\overline{\psi_\lambda(g)}$ (remember the
discussion in Remark \ref{BRP-CPT}) and a parity inversion
$\sigma_\mu\leftrightarrow \check\sigma_\mu=\sigma^\mu$.
Nevertheless, at this stage, a BRP$\leftrightarrow$CPT connection
inside the conformal group is just conjectural and it is still
premature to draw any physical conclusions based on it. It is not
either the main objective of this paper.

In this article we have considered a particular class of
representations (discrete series) of the conformal group, although
other possibilities could also be tackled. For example, we could
consider the new (vector and pseudo-vector) combinations
\[\tilde P_\mu\equiv\um(P_\mu+K_\mu),\;\;\tilde
K_\mu\equiv\um(P_\mu-K_\mu),\]
with new commutation relations:
\be\left[\tilde P_\mu,\tilde
K_\nu\right]=\eta_{\mu\nu}D,\;\left[\tilde P_\mu,\tilde
P_\nu\right]=M_{\mu\nu},\; \left[\tilde K_\mu,\tilde
K_\nu\right]=-M_{\mu\nu}.\label{noncomspace}\ee
Unlike in formulas (\ref{constrained3}) and (\ref{constrained4}),
the fact that now $\left[D,\tilde K_\mu\right]=-\tilde P_\mu$
precludes the imposition of $D^L, M^L_{\mu\nu}$ and $\tilde
K^L_\mu$ as a compatible set of constraints on wave functions.
Instead, we could impose
\[M^L_{\mu\nu}\psi=0, \;\;\tilde
K^L_\mu\psi=0\]
together with the Casimir (\ref{Casimir}) constraint
$C_2^L\psi=m_{00}^2\psi$, which leads to
\[
((D^L)^2+(\tilde P^L)^2)\psi=m_{00}^2\psi,\]
This equation could be seen as a \emph{generalized} Klein-Gordon
equation ($P^2\psi=m_0^2\psi$), with $D$ replacing $P_0$ as the
(proper) time generator and $m_{00}$ replacing the
Poincaré-invariant mass $m_0$, as a ``conformally-invariant mass''
(see e.g.\cite{Barutmass} for the formulation of other
conformally-invariant massive field equations of motion in
generalized Minkowski space). This means that Cauchy hypersurfaces
have dimension 4. In other words, the Poincaré time is a dynamical
variable, on an equal footing with position, the usual Poincaré
Hamiltonian $P_0$ suffering Heisenberg indeterminacy relations
too. Instead of the proper time (dilation) generator $D$, one
could also consider the new combination $\tilde P_0=(P_0+K_0)/2$
as the new Hamiltonian of our theory (see \cite{CMP55} for this
choice).

In a non-commutative geometry setting \cite{Madore}, the
non-vanishing commutators (\ref{noncomspace}), or those of the
position operators $X_\mu$ in (\ref{positionop}) giving spin
generators \cite{confposition1,confposition2}, can be seen as a
sign of the granularity (non-commutativity) of space-time in
conformal-invariant theories, along with the existence of a
minimal length or, equivalently, a maximal acceleration.

The appearance of a maximal acceleration inside the conformal
group will be manifest in analyzing the Unruh effect from a
group-theoretical perspective \cite{Unruhconf}. In a previous
paper \cite{conforme}, vacuum radiation in uniformly accelerated
frames was related to a spontaneous breakdown of the conformal
symmetry. In fact, in conformally-invariant quantum field theory,
one can find degenerated pseudo-vacua (which turn out to be
coherent states of conformal zero-modes) which are stable
(invariant) under Poincaré transformations but are excited under
accelerations and lead to a black-body spectrum. The same
spontaneous-symmetry-breaking mechanism applies to general
$U(N,M)$-invariant quantum field theories, where an interesting
connection between ``curvature and statistics'' has emerged
\cite{vacrad}. We hope this is just one of many interesting
physical phenomena that remain to be unravelled inside
conformal-invariant quantum field theory.

\section*{Acknowledgements}

Work partially supported by the Fundación Séneca (08814/PI/08),
Spanish MICINN (FIS2008-06078-C03-01) and Junta de Andaluc\'\i a
(FQM219). M.C. thanks the ``Universidad Politécnica de Cartagena''
and C.A.R.M.  for the award  ``Intensificación de la Actividad
Investigadora''. We all thank V. Aldaya for stimulating
discussions.


\begin{thebibliography}{99}
\bibitem{Coquereaux} R. Coquereaux and A. Jadczyk, Conformal theories, curved phase spaces,
relativistic wavelets and the geometry of complex domains, Rev.
Math. Phys. \textbf{2} (1990) 1-44

\bibitem{Odzijewicz} G. Jakimowicz and A. Odzijewicz, Quantum
complex Minkowski space, J. Geom. Phys. \textbf{56} (2006)
1576-1599

\bibitem{Penrose} R. Penrose, The twistor programme, Rep. Math. Phys. \textbf{12} (1977) 65-76.
\bibitem{Penrose2} Penrose, R., and W. Rindler, Spinors and Space-Time, Volume 2:
Spinor and Twistor Methods in Space-Time Geometry, Cambridge
University Press, 1986

\bibitem{Souriau} J.M. Souriau: Structure des systemes dynamiques, Dunod Paris
(1970)
\bibitem{Kirillov} A.A. Kirillov, Elements of the Theory of
Representations, Springer, Berlin, (1976).

\bibitem{Ruhl0}  W. R\"uhl, {Distributions on Minkowski space and their connection
with analytic representations of the conformal group}, Commun.
Math. Phys. \textbf{27} (1972) 53-86.

\bibitem{Ruhl1} W. R\"uhl, {Field Representations of the Conformal
Group with Continuous Mass Spectrum}, Commun. Math. Phys.
\textbf{30} (1973) 287-302.

\bibitem{Bal} A.P. Balachandran, G. Marmo, B-S. Skagerstan and A. Stern, Gauge Theories and
Fiber Bundles: Applications to Particle Dynamics, Lecture Notes in
Physics \textbf{188} (Springer-Verlag, Berlin, 1983)
\bibitem{Bal2} G. Alexanian, A.P. Balachandran, G. Immirzi and B. Ydri, Fuzzy $\mathbb
CP^2$, J. Geom. Phys. \textbf{42} (2002) arXiv:hep-th/0103023v2

\bibitem{hamilredu} V. Gerdt, R. Horan, A. Khvedelidze, M. Lavelle,
D. McMullan and Yu. Palii, On the Hamiltonian reduction of
geodesic motion on SU(3) to SU(3)/SU(2), J. Math. Phys.
\textbf{47} (2006) 1129 arXiv:hep-th/0511245v1

\bibitem{KaiserAP} G. Kaiser, {Quantized Fields in Complex
Spacetime}, Ann. Phys. \textbf{173} (1987)  338-354

\bibitem{Kaiser} G. Kaiser, {Quantum Physics, Relativity, and Complex Spacetime:
Towards a New Synthesis}, North-Holland, Amsterdam (1990)
arXiv:0910.0352v2 [math-ph]

\bibitem{Jadczyk} A. Jadczyk, Born's reciprocity in the conformal domain, in ``Spinors, Twistors, Clifford Algebras and
Quanfum Deformations'', Eds. Z. Oziewicz et al., Kluwer Academic
Publ., 1993, pp. 129-140.

\bibitem{Born1} M. Born, A suggestion for unifying quantum theory and relativity,
Proc. R. Soc. \textbf{A165} (1938) 291-302
\bibitem{Born2} M. Born, Reciprocity theory of elementary particles, Rev. Mod.
Phys. \textbf{21} (1949) 463-473

\bibitem{Castro0} C. Castro and M. Pavsic, Clifford Algebra of
Spacetime and the Conformal Group, Int. J. Theor. Phys.
\textbf{42} (2003) 1693-1705

\bibitem{Castro1} C. Castro, The Extended Relativity Theory in Born-Clifford
Phase Spaces with a Lower and Upper Length Scales and Clifford
Group Geometric Unification, Foundations of Physics \textbf{35}
(2005) 971-1041

\bibitem{Caianiello} E.R. Caianiello, Is there a maximal aceleration?,
Lettere al Nuovo Cimento \textbf{32} (1981)  65-70

\bibitem{Castro2} C. Castro, On the variable fine structure constant, strings
and maximal-acceleration phase space relativity, International
Journal of Modern Physics \textbf{A18} (2003) 5445-5473

\bibitem{Low1} S.G. Low, Canonically relativistic quantum mechanics:
representations of the unitary semidirect Heisenberg group $U(1,
3)\otimes_s H(1,3)$, J. Math. Phys. \textbf{38}  (1997) 2197-209


\bibitem{Low2} S.G. Low, Representations of the canonical group, (the semidirect product of the unitary and Weyl-Heisenberg groups), acting as a
dynamical group on noncommutative extended phase space, J. Phys.
\textbf{A35} (Math.\&Gen.) (2002) 5711-5729


\bibitem{Fulling} S.A. Fulling, Nonuniqueness of Canonical Field Quantization in Riemannian Space-Time, Phys. Rev. \textbf{D7}  (1973) 2850-2862
\bibitem{Davies} P.C.W. Davies, Scalar production in Schwarzschild and Rindler metrics, J. Phys. \textbf{A8} (1975)  609-616
\bibitem{Unruh} W. G. Unruh, Notes on black-hole evaporation, Phys. Rev. \textbf{D14} (1976) 870
 \bibitem{Hawking} S.W. Hawking, Black-hole explosions?, Nature \textbf{248} (1974) 30

\bibitem{Unruhconf}   M. Calixto, E. Pérez-Romero and V. Aldaya,  Group-Theoretical Revision of the Unruh Effect and
Maximal Acceleration, in preparation.

\bibitem{conforme} V. Aldaya, M. Calixto and J.M. Cerveró, Vacuum radiation and symmetry breaking in conformally
invariant quantum field theory, Commun. Math. Phys. \textbf{200}
(1999) 325-354

\bibitem{Kaiser1} G. Kaiser, {Space-time-scale Analysis of Electromagnetic
Waves}, Proceedings of the IEEE-SP International Symposium on
Time-Frequency and Time-Scale Analysis, Victoria, (1992) 209-212.

\bibitem{Kaiser2} G. Kaiser, {Wavelet Electrodynamics II: Atomic Composition of Electromagnetic
Waves}, Appl. Comput. Harmon. Anal. \textbf{1} (1994) 246-260.

\bibitem{EMSMTA} M. Calixto and E. Pérez-Romero, Extended MacMahon-Schwinger's Master Theorem
and Conformal Wavelets in Complex Minkowski Space,
arXiv:1002.3498v2

\bibitem{Hill} E. L. Hill, On accelerated coordinate systems in classical and
relativistic mechanics, Phys. Rev. {\bf 67} (1945) 358-363.

\bibitem{conforme-ac} T. Fulton, F.  Rohrlich and L.  Witten,
Physical consequences to a coordinate transformation to a
uniformly accelerationg frame, Nuovo Cimento \textbf{26} (1962) 652-671



\bibitem{Cervero1975}  L. J. Boya and J. M. Cerveró, Contact Transformations and Conformal Group. I. Relativistic Theory,
International Journal of Theoretical Physics \textbf{12} (1975)
47-54

\bibitem{Weyl} H. Weyl, ``Space, Time and Matter''.
Dover. NY. First Edition 1922.


\bibitem{Kastrup1} H.A. Kastrup, Gauge properties of the Minkowski space, Phys. Rev. {\bf 150},  1183 (1966).


\bibitem{expomap0} A. O. Barut, J. R. Zenitt and A Lauferts, The exponential map for the conformal group
SO(2,4), J. Phys.  \textbf{A27} (Math. Gen.) (1994) 5239-5250.
\bibitem{expomap1}
A. O. Barut, J. R. Zenitt and A Lauferts, The exponential map for
the unitary group SU(2,2), J. Phys.  \textbf{A27} (Math. Gen.)
(1994) 6799-6805.
\bibitem{conformecontract} V. Aldaya and J.A. de Azcárraga, Group manifold
analysis of the structure of relativistic quantum dynamics, Ann.
Phys. \textbf{165} (1985) 484-504
\bibitem{dilatatiempo} S. Fubini, A.J. Hanson and R. Jackiw, New approach
to field theory, Phys. Rev. \textbf{D7} (1973) 1732-1760
\bibitem{Ramirez} V. Aldaya, J. Navarro-Salas and A. Ramirez, Algebraic quantization on a group and nonabelian constraints,
 Commun. Math. Phys. \textbf{121}, 541 (1989)
\bibitem{McMullan} D. MacMullan and I. Tsutsui, On the emergence of gauge
structures and generalized spin when quantizing on a coset space, Ann. Phys.
\textbf{237} (1995) 269-321
\bibitem{FracHall} V. Aldaya, M. Calixto and J. Guerrero,  Algebraic Quantization, Good Operators
and Fractional Quantum Numbers, Commun. Math. Phys. \textbf{178},
399 (1996)
\bibitem{Landsmann} N.P. Landsman and N. Linden, The geometry of inequivalent quantizations,
Nucl. Phys. \textbf{B365}, 121 (1991)

\bibitem{Woodhouse} N. Woodhouse, Geometric Quantization, Oxford University Press
(1980)
\bibitem{GAQ} V. Aldaya and J.A. de Azc\'{a}rraga, Quantization as a consequence of the symmetry group: an approach to geometric quantization,
 J. Math. Phys. \textbf{23} (1982)
1297
\bibitem{higherpol} V. Aldaya, J. Guerrero and G. Marmo, Higher-order differential operators on a Lie
group and quantization, Int. J. Mod. Phys. \textbf{A12} (1997) 3

\bibitem{confposition1} Marc-Thierry Jaekel and Serge Reynaud, Space-time localisation with quantum
fields, Phys. Lett. \textbf{A220} (1996) 10-16

\bibitem{confposition2} Marc-Thierry Jaekel and Serge Reynaud, Conformal Symmetry and Quantum Relativity,
Found. Phys. \textbf{28} (1998) 439-456

\bibitem{acha} M. Calixto and J. Guerrero, {Wavelet  Transform on the
Circle and the Real Line: a Group-Theoretical Treatment}, Appl.
Comput. Harmon. Anal. \textbf{21} (2006) 204-229.

\bibitem{Perelomov} A. Perelomov, Generalized Coherent States and Their
Aplications, Springer-Verlag (1986)
\bibitem{Klauder} J.R. Klauder and
Bo-Sture Skagerstam, Coherent States: Applications in Physics and
Mathematical Physics, World Scientific (1985)
\bibitem{Mulak} W. Mulak, Quantum $SU(2,2)$-harmonic oscillator, Rep. Math. Phys. \textbf{34} (1994) 143-149


\bibitem{Castro3} C. Castro,
Born's Reciprocal General Relativity Theory and Complex
Non-Abelian Gravity as Gauge Theory of the Quaplectic Group: a
Novel Path to Quantum Gravity, Int. J. Mod. Phys. \textbf{A23}
 (2008) 1487-1506

\bibitem{conforme-maxac} W.R. Wood, G. Papini and Y.Q. Cai,
Conformal transformations and maximal acceleration, Il Nuovo
Cimento \textbf{104} (1989) 653-663

\bibitem{Barutmass} A.O. Barut and R.B. Haugen, Theory of the Conformally Invariant
Mass, Ann. Phys. \textbf{71}  (1972) 519-541

\bibitem{CMP55} G. Mack, All Unitary Ray Representations of the Conformal Group
SU(2,2) with Positive Energy, Commun. Math. Phys. \textbf{55} (1977) 1-28


\bibitem{Madore} J. Madore, {An Introduction to Noncommutative
Differential Geometry and its Physical Applications}, 2nd ed.,
London Mathematical Society Lecture Note Series, \textbf{257}
Cambridge Univ. Press. 1999

\bibitem{vacrad} M. Calixto and V. Aldaya, Thermal Vacuum Radiation in Spontaneously
Broken Second-Quantized Theories on Curved Phase Spaces of
Constant Curvature, Int. J. Geom. Meth. Mod. Phys. \textbf{6}
(2009) 513-531



\end{thebibliography}
\end{document}